\documentclass{aa}  
\usepackage{hyperref}
\usepackage[normalem]{ulem}
\hypersetup{
   colorlinks=true,
   linkcolor=blue,
   filecolor=blue, 
   citecolor=blue, 
   urlcolor=blue, 
   pdftitle={Overleaf Example},
   pdfpagemode=FullScreen,
   }
\usepackage{textcomp}
\usepackage{dblfloatfix}
\usepackage{graphicx}
\usepackage{xspace}
\usepackage{tabularx}
\usepackage{booktabs}
\usepackage{xcolor}
\usepackage{subfig}
\usepackage{txfonts}
\newcommand{\hh}{$\rm H_2$\xspace}
\newcommand{\kms}{$\rm km \, s^{-1}$\xspace}
\newcommand{\co}{$\rm C^\mathrm{18}O$\xspace}
\newcommand{\cco}{$\rm ^{13}CO$\xspace}
\newcommand{\cm}{$\rm \mathrm{cm}^{-2}$\xspace}

\begin{document} 

   \title{Unveiling the role of magnetic fields in a filament accreting onto a young protocluster}

   \author{Farideh S. Tabatabaei
          \inst{1}
          \and
          Elena Redaelli\inst{1}
           \and
           Daniele Galli\inst{2}
           \and
            Paola Caselli\inst{1}
           \and
          Gabriel A. P. Franco\inst{3}
          \and
          Ana Duarte-Cabral\inst{4}
          \and
           Marco Padovani\inst{2}
          }

   \institute{Centre for Astrochemical Studies, Max-Planck-Institut für extraterrestrische Physik, Gießenbachstraße1, 85749 Garching bei München, Germany
   \and
   INAF-Osservatorio Astrofisico di Arcetri, Largo E. Fermi 5, 50125 Firenze, Italy
   \and
   Departamento de Física—ICEx—UFMG, Caixa Postal 702, 30.123-970 Belo Horizonte, Brazil
   \and
    School of Physics \& Astronomy, Cardiff University, Queen’s Building, The Parade, Cardiff, CF24 3AA, UK
             }

   \date{Received XXX; accepted XXX}


  \abstract
 { To develop a more comprehensive picture of star formation, it is essential to understand the physical relationship between dense cores and the filaments embedding them. There is evidence that magnetic fields play a crucial role in this context.  }
   {We aim to understand how magnetic fields influence the properties and kinematics of an isolated filament located east of the Barnard 59 clump, within the Pipe Nebula.}
   {We used near-infrared polarization observations to determine the magnetic field configuration, and we applied the Davis–Chandrasekhar–Fermi method to infer the magnetic field strength in the plane of the sky.
Furthermore, we used complementary data from the \textit{James Clerk Maxwell} Submillimetre Telescope of \co and the \cco $J=3-2$ transition to determine the filament's kinematics. Finally, we modeled the radial density profile of the filament with polytropic cylindrical models.}
   { Our results indicate that the filament is stable to radial collapse and is radially supported by agents other than thermal pressure. In addition, based on previous observations of emission lines on this source (\citealt{duarte12}), we suggest that gas is flowing toward the hub, while \co (3-2) nonthermal motions indicate that the cloud is in a quiescent state. }
   {}


   \maketitle
%

\section{Introduction}

 \begin{figure*}
\sidecaption
  \includegraphics[width=12cm]{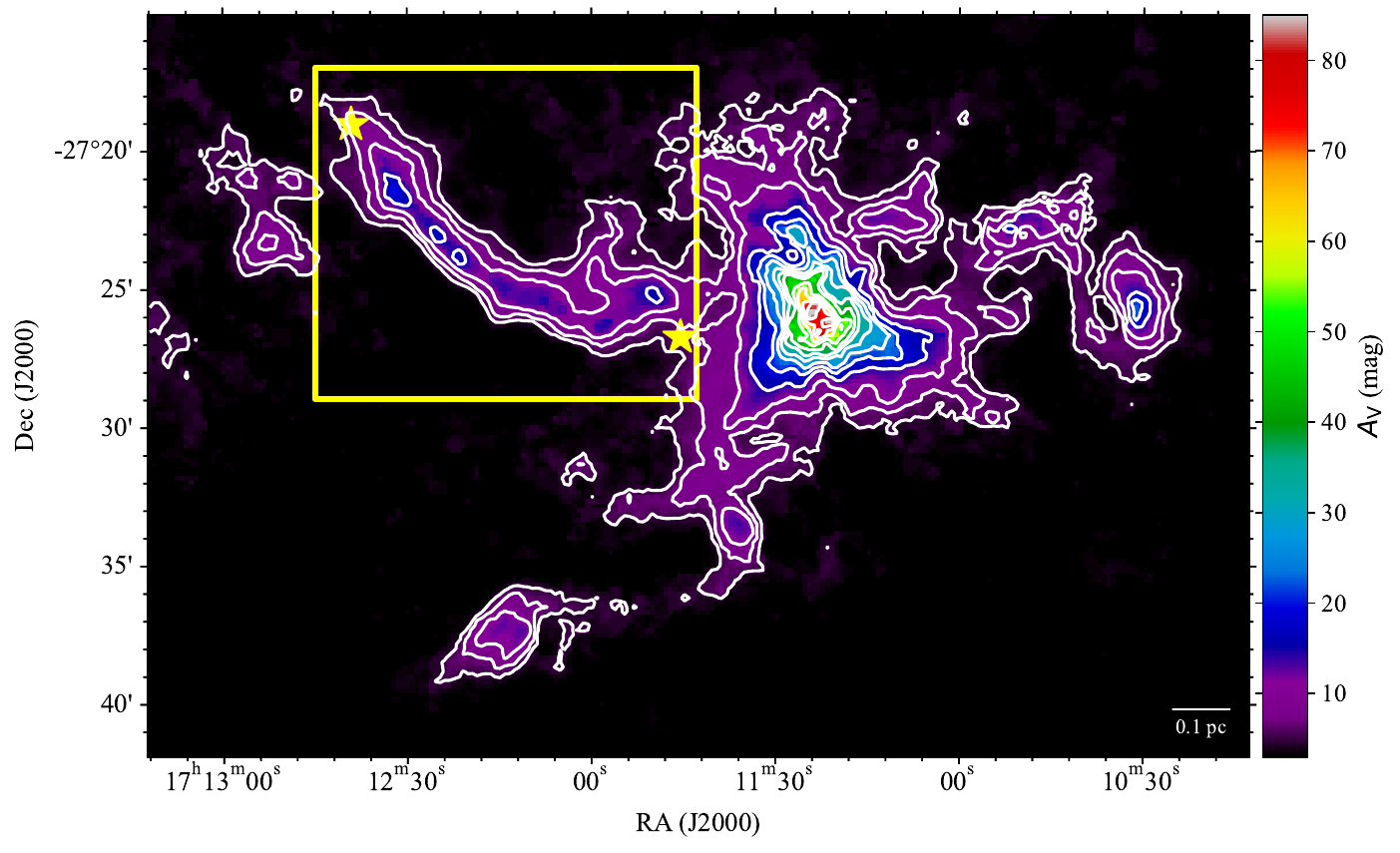}
     \caption{Dust extinction map of the B59 region at a spatial resolution of 20$^{\prime \prime}$ (\citealt{roman10}). The solid white line contours represent levels of visual extinction at A$_\mathrm{V} = [5, 7, 10, 15, 20, 25, 30, 35, 40, 50, 60, 70, 80, 90$] mag. The yellow square marks the filament studied in this work. Two yellow stars indicate the approximate length of the filament. The scale bar is at the bottom right side of the panel.}
     \label{B59}
\end{figure*}

%
%

The process of low-mass star formation is initiated by the fragmentation of molecular clouds into cold, dense cores. These cores are usually identified by subsonic gas motions, in contrast with the supersonic turbulence of their surrounding clouds (\citealt{goodman98}). The formation of a protostar is determined by a competition between gravity, thermal pressure, external pressure, and magnetic field strength. The precise role of the magnetic field (B-field) in the star formation process is still controversial, highlighting the importance of studying the configuration of B-field lines in regions where stars are formed.

According to \textit{Planck} data analysis, clouds are generally magnetized (\citealt{planck16}). The B-field tends to be aligned parallel to the density structures for visual extinction ($A_V$) values less than or equal to $2-3.5$ mag. However, beyond this A$_V$ threshold, the B-field orientation becomes perpendicular to the density structures (\citealt{soler17}; \citealt{planck16}). 
Studies show that the B-field may take on a pinched morphology perpendicular to the long axis of the filament (e.g., \citealt{tomisaka15}; \citealt{burge16}). For example, higher-resolution observations strongly indicate a significant pinch of the B-field inside the filament of NGC 1333 (\citealt{yasuo21}).

In the low-mass star formation regime, pre-stellar cores are the dense cores where star formation is about to take place (i.e., prior to the formation of a protostar in the center), and they are embedded in larger structures that often exhibit filamentary shapes. The origin of these filamentary structures is still debated.
Two main hypotheses have been proposed to explain their formation. The first suggests that filaments are formed due to compressions in turbulent clouds (\citealt{padoan01} and \citealt{auddy16}). The second proposes that filaments arise from the global gravitational collapse of a molecular cloud as a whole, and consequently, filaments accrete material from their parent clouds and feed it into the bottom of the potential well (\citealt{heit-a}, \citealt{heit-b}, and \citealt{henn13}).

Observational evidence suggests that there are organized gas flows along filaments toward the cores in both high-mass and low-mass star-forming regions (see \citealt{henshaw13} for an example of filamentary structure in high-mass star-forming regions and \citealt{hacar13} for low-mass star-forming regions). This is supported by studies by \cite{kirk13} of gas flows along the southern filament of the Serpens South embedded cluster (see also \citealt{peretto14}).
To gain a better understanding of star formation, it is crucial to comprehend the physical relationship between cores and the filaments they are associated with.

The Pipe Nebula, located at a distance of $d_\mathrm{cloud} = 163$ pc (\citealt{dzib}), is one of the closest known star-forming regions. The star formation rate within the Pipe cloud is notably low (only $\sim$0.06\% efficiency; \citealt{forb09}). \cite{lambordi2006} identified 159 dense cores across the Pipe Nebula, each with estimated masses ranging from 0.5 to 28 $ \mathrm{M}_\odot$. However, it appears that the only area within this cloud where star formation is currently taking place is limited to its northwestern extreme, known as Barnard 59 (hereafter B59; \citealt{brooke07}; \citealt{forb09}; \citealt{forb10}). B59 contains a young protocluster with approximately 20 young stellar objects (\citealt{brooke07}, \citealt{redaelli2017}). 

\cite{alves08} and \cite{franco10} proposed that the magnetic properties of the Pipe Nebula may be responsible for the fact that only B59 has active star formation.
Studies of the morphology of the B-field with optical polarimetric observations revealed the filamentary structure to be threaded by a B-field, aligned perpendicular to the cloud’s axis, that is possibly preventing or slowing down its global collapse (\citealt{alves08}; \citealt{franco10}).

As \cite{peretto12} suggested, B59 is located in the center of what resembles a hub-like clump with apparently converging filaments. Nevertheless, from the kinematics of the region, \cite{duarte12} determined that a significant fraction of the filaments around the B59 clump are shaped by the outflows of the forming protocluster pushing the cloud material, rather than by infalling filaments of gas. The one clear exception is the filament to the northeast of the clump, which is the target of this work. 
This filament, although situated in the immediate vicinity of the B59 star-forming clump, is relatively isolated and seems to be extremely quiescent (Fig. \ref{B59}).
As such, it is a good candidate for our modeling (see Sect. \ref{sec:model}).

This paper is organized as follows. In Sect. \ref{sec-obser} we present the data collected from different telescopes and describe how we processed them. In Sect. \ref{sec-kin} we explain how we analyzed the data and discuss our results in terms of the kinematic properties of the filament. In Sect. \ref{sec-pola} we discuss the polarization and B-field properties through the filament, and lastly, we present our modeling of a hydrostatic filamentary cloud based on our observational parameters of the filament within B59 in Sect. \ref{sec:model}. Our main findings and conclusions are summarized in Sect. \ref{sec-concl}.

\section{Observations}
\label{sec-obser}

\subsection{Near-infrared observations}
The polarimetric data were obtained at Observat\'orio do Pico dos Dias/Laborat\'orio Nacional de Astrof\'isica (OPD/LNA, Brazil) using IAGPOL, the IAG imaging polarimeter mounted on the 1.6 m telescope. Observation runs were completed in June 2013.
A special infrared (IR) detector was used to collect these data for polarimetric measurements. A precise description of the polarimeter is provided in \cite{maglh96}.
We obtained linear polarimetry in the H band using deep imaging for three fields, each with a 4$^\prime$ by 4$^\prime$ field of view.
The images were obtained with an IR camera, CamIV, which has a HAWAII detector of 1024 by 1024 pixels and 18.5 $\mu$m/pixel, resulting in a plate scale of 0.25$^{\prime\prime}$/pixel. 
For each of the eight waveplate positions separated by 22.5$^{\circ}$, sixty dithered images were obtained following a five-dot pattern (12 by 5 positions).
Each image was exposed for 10 seconds, adding up to 600 seconds per wave-plate position.
Observations of polarized standard stars were used to determine the reference direction of the polarizer.

\begin{figure*}[!ht]
\centering
\includegraphics[width=\hsize]{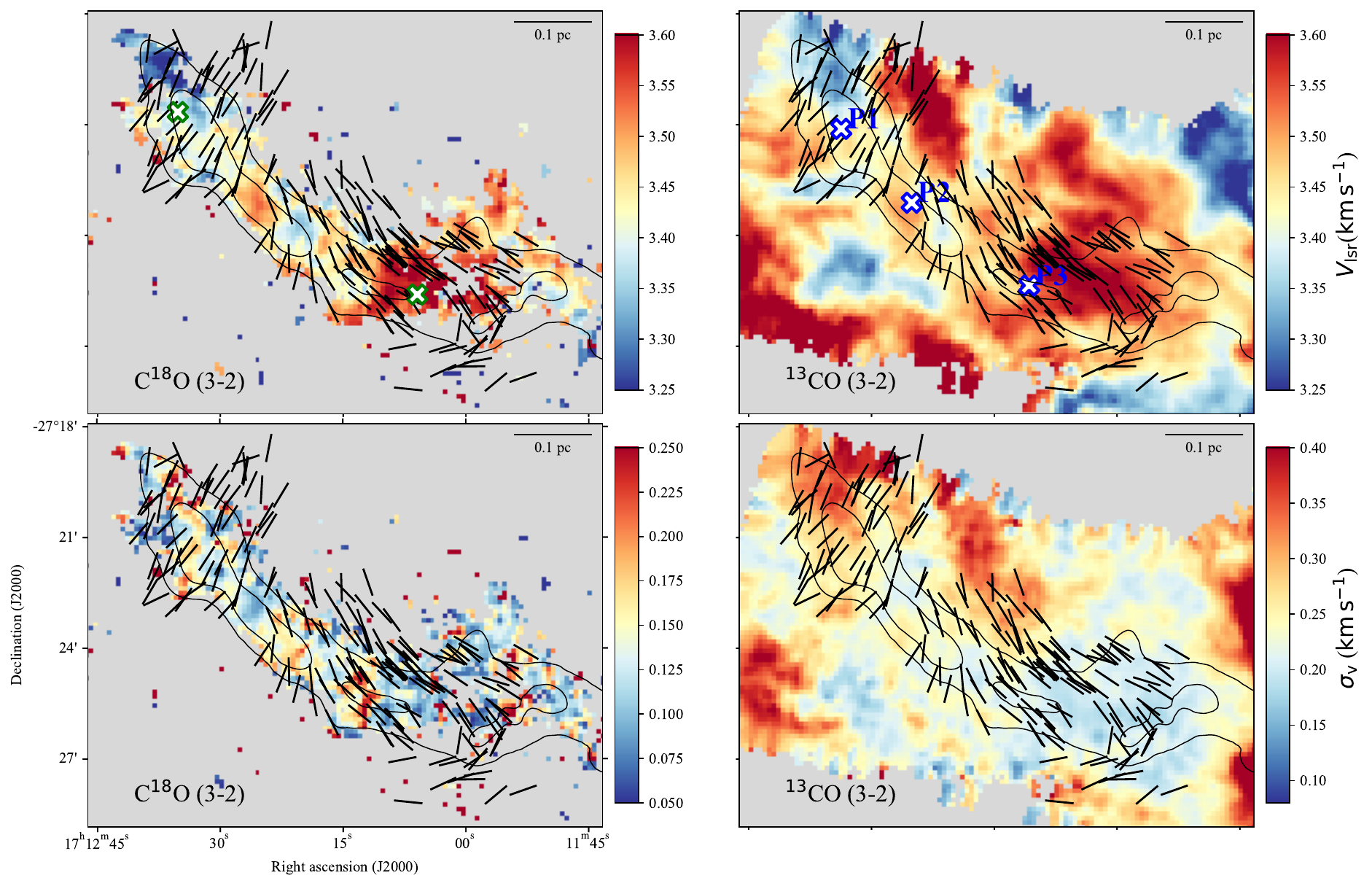}
\caption{ Maps of kinematic parameters. Top panels: Centroid velocity maps of \cco (3-2) and \co (3-2) obtained from the Gaussian fitting procedure described in Sect. \ref{sec-kin}, Bottom panels: Velocity dispersion maps of \cco and \co. Contours show the column density of \hh with levels [4,6] $\times 10^{22}$ \cm, as derived from $\emph{Herschel}$ data (\citealt{peretto12}). The black segments represent the polarization angles in the NIR band. The blue crosses in the top-right panel present the locations of spectra in Fig. \ref{spec}. The green crosses represent the locations where the velocity gradient is estimated. Scale bars are at the top right of each panel. }
\label{kinematics}
\end{figure*}

From the Stokes $Q$ and $U$, we derived degree to the linear polarization, 
 $P= \sqrt{ Q^2 + U^2}$ and the polarization angle, $\theta_\mathrm{p} = 0.5  \mathrm{tan}^{-1} (U/Q)$.
Assuming that the polarization is produced by magnetically aligned dust grains, in all the figures, we show polarization segments that outline the direction of the B-field. Since near-infrared (NIR) polarization is produced by dichroic extinction, polarization segments give the B-field direction directly.
The data are filtered by a $ P/\sigma_\mathrm{P} \geq 3$ where correspond to an uncertainty in $\theta$ ($\sigma_\theta < 10^{\circ} $) and $\sigma_\mathrm{P}$ represents the standard deviation of the measurement. The data are length-scaled by the polarization degree, with the longest vectors with $\sim$ 3\% polarization and a mean value of $\sim$1\% (segments in Fig. \ref{kinematics} and Fig. \ref{extinction}).


Our analysis of gas column density and dust temperature is based on data from the Gould Belt Survey obtained with the $\emph{Herschel}$ Space Telescope (SPIRE/PACS instrument; \citealt{peretto12}, \citealt{roy14}). 
We utilized the visual extinction map created from a deep near-infrared (NIR) imaging survey conducted with various telescopes, including the New Technology Telescope (NTT), the Very Large Telescope (VLT), and the 3.5-meter telescope at the Centro Astron\'omico Hispano Alem\'an (CAHA) as well as 2MASS data (\citealt{roman10}, Fig. \ref{B59}).

\subsection{\textit{James Clerk Maxwell} Telescope}

The molecular line data used in this work are presented in \cite{duarte12}, where a detailed description of the observations and data reduction can be found. In brief, here we use the maps of \cco (3-2) and \co (3-2) at 330.6 and 329.3 GHz, respectively, covering the entire B59 star-forming region ( $\sim 0.11$ deg$^2$). The data were obtained from the \textit{James Clerk Maxwell} Telescope (JCMT) using HARP (\citealt{buckle}) in May and June 2010. This dataset contains an rms noise level of $0.22 $ K (in $T_\mathrm{A}^\star$) with 0.25 \kms channels. The main beam efficiency of the telescope ($\eta_\mathrm{mb} = 0.66$) was taken into account for all data (\citealt{buckle}; \citealt{curtis10})

%




\section{Kinematical analysis of the filament}
\label{sec-kin}

To interpret the gas kinematic behavior in the filament, we estimated the centroid velocity ($V_\mathrm{lsr}$) and velocity dispersion ($\sigma_\mathrm{v}$) by fitting a single Gaussian profile to the \co and \cco line emission, using the Python package $pyspeckit$ (\citealt{ginsburg11}).  
All fitted spectra have residuals of less than 2 $\times$ rms and widths of less than 0.35 \kms. 
Almost all spectra with S/N > 3.5 for \co and S/N > 4.5 for \cco are well fitted using only one Gaussian, as seen in Fig. \ref{spec}, as proved by the low residuals (see Appendix \ref{appen1}). As a result, and for simplicity, we adopted the single-velocity component fit for the whole filament.
The best-fit parameters of $V_\mathrm{lsr}$ and $\sigma_\mathrm{v}$ are shown in Fig. \ref{kinematics}. 
The $V_\mathrm{lsr}$ map of \co shows the velocity gradient of gas toward the location of the hub, ranging from 3.3 \kms to 3.6 \kms. To estimate the velocity gradient, we determined an approximate estimate of two positions at a distance of $\Delta R$ $\sim$ 0.4 pc (see the green crosses in Fig. \ref{kinematics}). Therefore, the velocity gradient for \co is $\nabla V = \Delta V / \Delta R = 0.69 \pm 0.01 $ km s$^{-1}$ pc$^{-1}$. This value is comparable with that of a smaller filament, $\sim$ 0.1 pc, in Lupus I. In comparison, \cite{taba} report $\nabla V = 0.9 \pm 0.2$ km s$^{-1}$ pc$^{-1}$ for a filament associated with a low-mass Class 0 stellar object IRAS15398-3359.


\begin{figure}[!ht]
   \centering
   \includegraphics[width=\hsize]{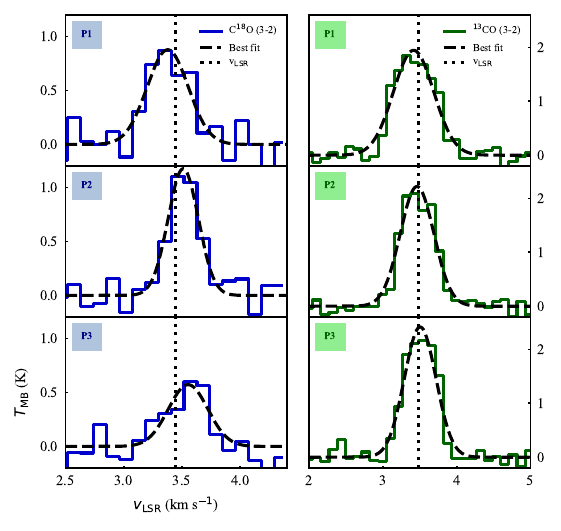}
      \caption{ \co (3-2) and \cco (3-2) spectra at three selected locations within the filament (locations are shown with blue crosses in Fig. \ref{kinematics}). \co (3-2) spectra are in the left column, and \cco (3-2) spectra are in the right column. Blue and green lines show the observed spectra for the \co and \cco, respectively, whereas dashed black curves show the best-fit Gaussian for each. As displayed by the vertical dotted lines, \co (3-2) and \cco (3-2) have average V$_\mathrm{lsr}$ of 3.4 and 3.5 \kms, respectively.
              }
         \label{spec}
\end{figure}

\subsection{ Velocity dispersion and nonthermal motions}

The total observed broadening of any optically thin line profile is the sum of the contributions of thermal and nonthermal motions.
We can obtain some insight into the gas kinematics properties by comparing the thermal sound speed with the Gaussian velocity dispersion.

To estimate nonthermal velocity dispersion, we subtracted the thermal velocity dispersion from the observed velocity dispersion in quadrature:

\begin{equation}
\begin{split}
\label{non-thermal}
\sigma_\mathrm{nt} = \sqrt{ \sigma_\mathrm{obs}^2 - \sigma_\mathrm{th}^2},
\end{split}
\end{equation}
where $\sigma_\mathrm{obs}$ is equal to $\sigma_\mathrm{v}$ (see Sect. \ref{sec-kin}). We used the mean value of $\sigma_\mathrm{v}$ for the filament area as defined by $A_\mathrm{V} > 8$ mag.

The thermal velocity dispersion is

\begin{equation}
\begin{split}
\label{thermal}
\sigma_\mathrm{th} = \sqrt{ \frac{k_\mathrm{B} T_\mathrm{kin} }{ m_\mathrm{obs}}},
\end{split}
\end{equation}
where $T_\mathrm{kin}$ is the kinetic temperature of the gas, $k_\mathrm{B}$ is Boltzmann constant and $m_\mathrm{obs}$ is the mass of observed molecule. 
Hence, the effect of $\sigma_\mathrm{nt}$ is quantified by the turbulent Mach number, which is calculated as the ratio of nonthermal dispersion to isothermal sound speed $\mathcal{M} = \sigma_\mathrm{nt} / c_\mathrm{s}$. The isothermal sound speed was obtained from
\begin{equation}
\begin{split}
\label{sound}
c_\mathrm{s} = \sqrt{ \frac{k_\mathrm{B} T_\mathrm{kin}}{\mu m_\mathrm{H}}},
\end{split}
\end{equation}
where $\mu = 2.37$ (\citealt{kauffmann08}) is the mean molecular weight per free particle and $m_\mathrm{H}$ is the mass of the hydrogen atom.
We used the dust temperature, $T_\mathrm{d}$, as a proxy for the gas kinetic temperature. We obtain $T_\mathrm{d}$ = 14 K from the $\emph{Herschel}$ map. 
We obtain $c_\mathrm{s} = 0.2$ \kms and the corresponding $\sigma_\mathrm{nt}$ are $0.12 \pm 0.02$ \kms and $0.22 \pm 0.01$ \kms for \co and \cco, respectively.
For this filament, \cite{duarte12} estimated an excitation temperature of $9$ K for \cco. The thermal velocity dispersion, $\sigma_\mathrm{th}$, is reduced by $\sim$ 20$\%$ when $T_\mathrm{kin} = 9$ K is used in Eq.\ref{thermal}.
According to the \co data, nonthermal motions are subsonic ($\mathcal{M} = 0.54 < 1$). This suggests that the filamentary cloud is in a quiescent state where the nonthermal gas motions are smaller than the sound speed.
We highlight that for this filament, for the \cco emission, we obtain $\mathcal{M} = 1.0$, implying that the gas motions are transonic. The difference between the Mach number of these two molecules comes from the difference in their velocity dispersion. The emission of \cco traces the lower-density gas, which is more turbulent and likely to be affected by larger optical depth, causing more line broadening than for the \co lines.


 \begin{figure*}
\sidecaption
  \includegraphics[width=12cm]{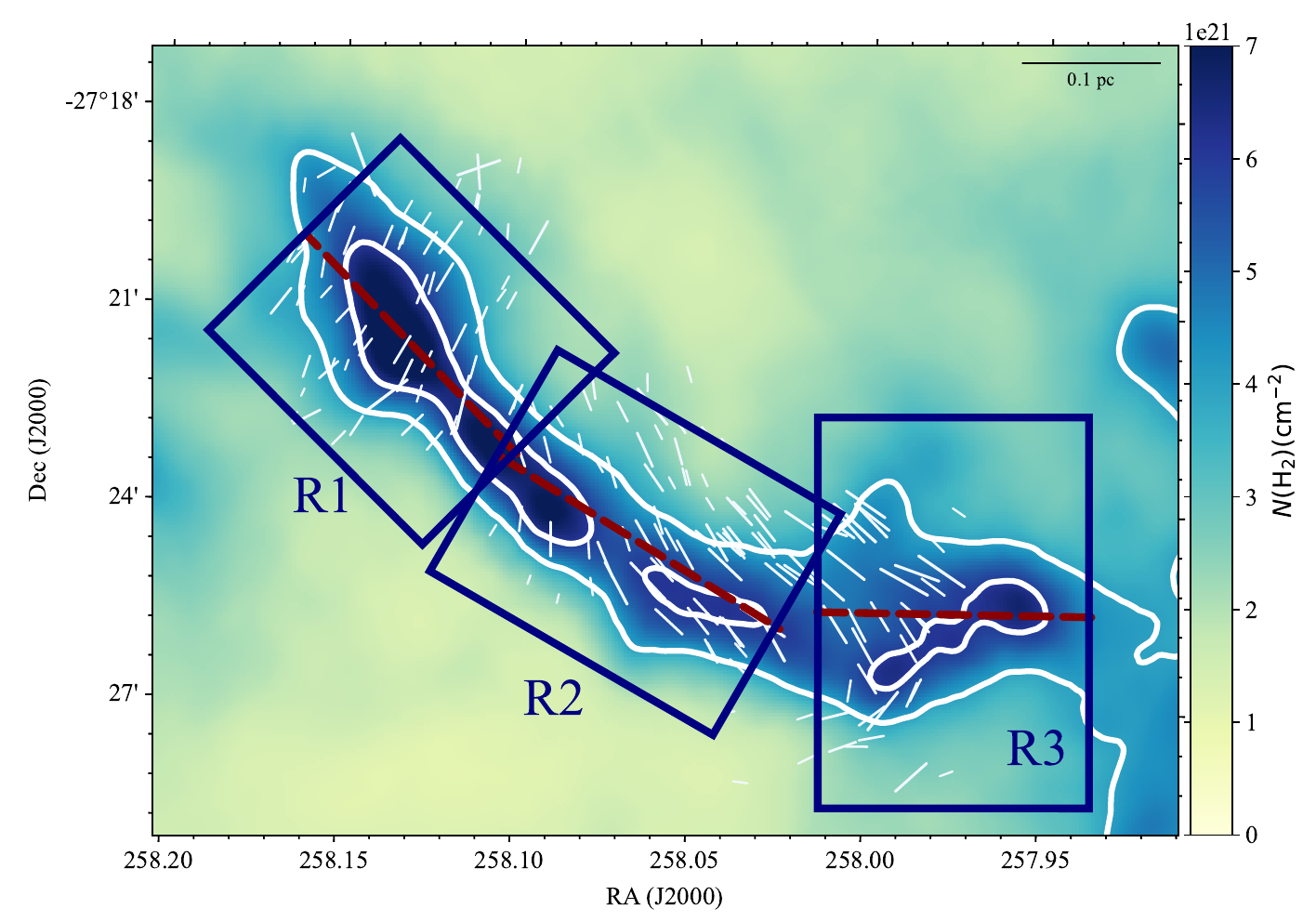}
     \caption{The column density map of the filament in B59 (from $\emph{Herschel}$ data) overlaid with the polarization segments from Pico dos Dias Observatory. The blue rectangles divide the filament into three regions labeled R1, R2, and R3. The red dashed lines represent the axis of each region. Contours show column density of \hh with levels: [4,6] $\times 10^{22}$ \cm. The scale bar is on the top right side of the panel.}
     \label{extinction}
\end{figure*}
%


\subsection{ Mass per unit length}
\label{sec:mass}
From a theoretical point of view, a self-gravitating isothermal cylinder has an infinite radius but a finite mass per unit length. The critical value of this parameter at 10 K is 
\begin{equation}
 \left(\frac{M}{L}\right)_\mathrm{cr} = \frac{2c_\mathrm{s}^2}{G} = 16\:  \left(\frac{T}{10 \mathrm{K}}\right) \: \mathrm{M}_{\odot} \: \mathrm{pc^{-1} },
\end{equation} 
where $G$ is the gravitational constant  $M_{\odot}$ and is the mass of the sun. 
If $\left(\frac{M}{L}\right)_\mathrm{fil} > \left(\frac{M}{L}\right)_\mathrm{cr}$, the filament collapses radially.
To estimate the filament mass, we used

\begin{equation}
 M_\mathrm{fil} = A \times \sum_{i=1}^n  \left(\frac{N(\mathrm{H}_2)}{\mathrm{cm}^{-2}}\right)_i  \left(\frac{\theta}{''}\right)^2 \left(\frac{d_\mathrm{cloud}}{\mathrm{pc}}\right)^2   \: \mu_\mathrm{H_2} \: m_\mathrm{H} \: M_{\odot},
\end{equation} 
where $\sum_{i=1}^n (N(\mathrm{H}_2)_i)$ adds the contribution of $n$ pixels within the area of the filament selected with $A_\mathrm{V} > 8$ mag, $m_\mathrm{H} = 1.67 \times 10^{-24}$ g is the hydrogen mass, $\mu_\mathrm{H_2}  = 2.8$ (from \citealt{kauffmann08}) is the gas mean molecular weight per hydrogen molecule, $d_\mathrm{cloud}$ is the distance to the cloud, $A = 1.13 \times 10^{-7}$ is a numerical factor to adjust units, and $\theta$ is the pixel size. In our case, $d_{\rm cloud} = 163$ pc and $\theta = 7.28 ''$. 
We obtained the gas column density of the filament, $N(\mathrm{H}_2)$, in two ways. First, we used the visual extinction, $A_\mathrm{V}$, a parameter that can be converted into molecular hydrogen column density using \cite{bohlin}'s relation, $N(\mathrm{H}_2) = 9.4 \times 10^{20} \times  A_\mathrm{V} $ \cm mag$^{-1}$. We also obtained $N(\mathrm{H}_2)$ from the $\emph{Herschel}$ map for the same exact area of the filament that was selected for the extinction map in the previous method. As a result, the mass calculated based on $\emph{Herschel}$ column density\footnotetext[1]{The column density map was constructed via pixel-by-pixel gray-body fitting from 160 $\mu$m to 500 $\mu$m, fixing the dust opacity such that $\kappa_\lambda = 0.1 \times \left(\frac{\lambda}{300 \mu m}\right)^{-2} \mathrm{cm^2 g^{-1}}$; for more details, see \cite{roy14}.}\footnotemark[1], $M_\mathrm{Herschel}$, and the mass calculated from the extinction map, $M_\mathrm{Ext}$, are 6.7 M$_{\odot}$ and 12.2 M$_{\odot}$, respectively. The resultant uncertainty is $\sim$ 42\% (\citealt{taba}, \citealt{roy14}, \citealt{dzib}) and $\sim$ 11.7\% (\citealt{roman10}) for $M_\mathrm{Herschel}$ and $M_\mathrm{Ext}$, respectively. The two values are consistent at $3 \sigma$ level.
We estimate $ L \, \sim $ 0.7 pc  for the filament length, which corresponds to the distance between the two yellow stars in Fig. \ref{B59}. We adopted both $M \, \sim $ 6.7 M$_{\odot}$ and 12.2 M$_{\odot}$ of the filament to obtain the observational value of the mass per unit length. 
Table \ref{Parameters3} shows the critical and observation values for the whole filament. 

In Sect. \ref{sec:model} we discuss the stability of the filament based on the cylindrical polytropic hydrostatic models by \citet[][hereafter TCa]{toci1}, and \citet[][hereafter TCb]{toci2}.
We also calculated the mass per unit length for different polytropic indexes of the cylindrical model. The mass per unit length is lower than the critical value for the whole filament in all the analyses (see Table \ref{Parameters3}), indicating that the filament is stable.

\begin{table*}[ht]
\caption{ Values of mass per unit length for the filament.} 
\centering 
\renewcommand{\arraystretch}{1.1}
\begin{tabular}{c c c c | c c c c c} 
\hline \hline
\toprule
  &  cr  & obs$_\mathrm{Herschel}^{(a)}$ & obs$_\mathrm{Ext}^{(b)}$ & Region & $n^{(c)}=-1$ & $n=-2$ & $n=-3$ & $n=-4$ \\  
\midrule
 $M/L$  ($ \mathrm{M}_{\odot}$/pc)  & 22.4  & 9.6 & 17.4 & R1 & 17.8 & 9.8 & 6.3 & 4.4  \\   
 $M/L$  ($ \mathrm{M}_{\odot}$/pc)  & 22.4  & 9.6 & 17.4 & R2 & 15.1 & 8.7 & 5.7 & 4.0  \\   

\hline \\ 
\end{tabular}
\label{Parameters3} 
\\
\footnotesize \textbf{Note}: The columns labeled cr, obs$_\mathrm{Herschel}$, and obs$_\mathrm{Ext}$ contain values corresponding to the whole filament region. $^{(a)}$ obs$_\mathrm{Herschel}$ is the mass per unit length value obtained from mass of filament calculated from  $\emph{Herschel}$ $N(\mathrm{H}_2)$ and $^{(b)}$ obs$_\mathrm{Ext}$ is from the mass of filament calculated with $A_\mathrm{V}$ (see Sect. \ref{sec:mass}).  $^{(c)}$ $n$ is the polytropic index; see Sect. \ref{sec:model}.
\end{table*}


\begin{figure*}[h!]
\centering

\includegraphics[width=0.48\textwidth]{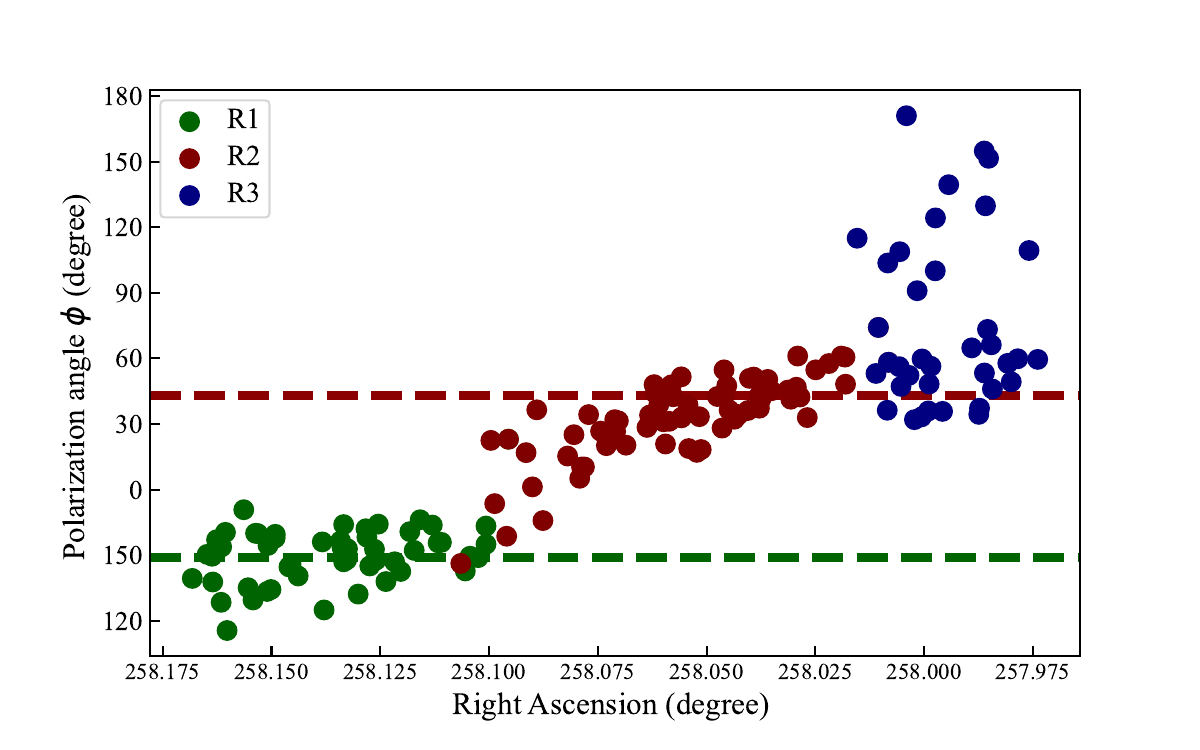}
\hfill
\includegraphics[width=0.48\textwidth]{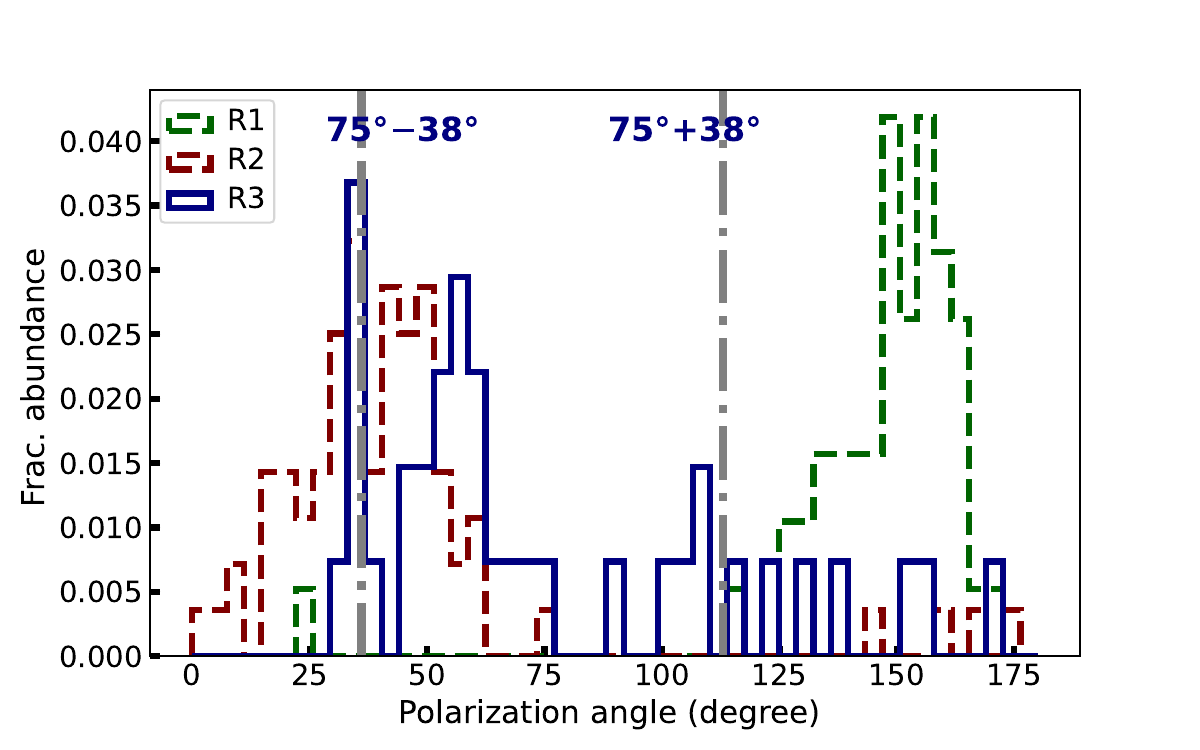}
\hfill    
        
\caption{ Distribution of polarization angles. Left panel: Distribution of polarization angles as a function of the right ascension of the field. Dashed green and red lines show the mean polarization angle values for R1 and R2, 147\textdegree, and 43\textdegree, respectively. Right panel: Histogram of the distribution of magnetic polarization angles. Green, red, and blue data points refer to R1, R2, and R3, respectively. Gray dash-dot lines illustrate the 
 1$\sigma$ interval around the mean for region R3.   }
     \label{distribution}
\end{figure*}

\section{Polarization and magnetic field properties of the filament}
\label{sec-pola}

We divided the filament area into three regions, R1, R2, and R3, as shown in Fig. \ref{extinction}. Although they belong to the same filament, each region has distinct polarimetric characteristics. Our method involves cutting the filament into three parts in which the B-field is approximately uniform in direction, as well as doing the cuts perpendicular to the filament's spine for each region (see the dashed red lines in Fig. \ref{extinction}). It is worth noting that the angles of the spine for the R1, R2, and R3 regions are precisely 45\textdegree, 60\textdegree, and 90\textdegree, respectively, from the north to the east. This allowed us to approximate the three regions as cylinders to which we can apply the model described in Sect. \ref{sec:model}.  
A detailed analysis of the distribution of the observed polarization angle as a function of Right Ascension over the core filament is presented in Fig.\ref{distribution}. 
The distribution of polarization angles is rather complex.
In region R1, we see a distribution of polarization angles ranging from $\sim$ 120\textdegree \ to 170\textdegree, which tends to concentrate around 149\textdegree $\pm$ 20\textdegree  (mean position angle $\pm$ standard deviation, dashed green line). All angles are measured from north to east.
In region R2 in Fig. \ref{distribution}, we can see that polarization angles range from $\sim$ -30\textdegree \, to 60\textdegree \, with a mean value of 34\textdegree $\pm 19$\textdegree (dashed red line).
In the small region southeast of the filament, region R3 (blue dots in Fig.\ref{distribution}), we obtain a wide range of polarization angles, 30\textdegree  \, to 170\textdegree \, with a mean value of 75\textdegree$\pm$ 38\textdegree. We present the 1$\sigma$ interval around the mean as two vertical gray lines in the right panel of Fig. \ref{distribution}. 
This analysis shows two sharp changes in the B-field orientation: the first change from R1 to R2 and another one inside of R3, the second presenting a U-shaped B-field where the filament is close to connecting to the core region.
These properties are more clearly illustrated in the histograms shown in Fig.\ref{distribution}, where two distributions of polarization angles are visible as two peaks. In the left panel of Fig. \ref{distribution} a curved structure can be observed from the R2 region to the R3 region, where the position angle increases as the RA decreases. The curvy structure of the filament likely causes it and tracks the change in the direction of the filament. 
The change in the B-field orientation from perpendicular to the filament spine in the eastern region to parallel toward the west, where this dense filament connects to the hub, suggests a coupled evolution of the B-field and the filament. 
The reorganization of the B-field along filaments could be caused by local velocity flows of matter in-falling toward the hubs, where the B-field is dragged by gravity and flows along the filaments, which is also observed in magnetohydrodynamic  (MHD) simulations (e.g., \citealt{gomez18}). \cite{pillai20} studied the Serpens hub-filament with polarimetric data from NIR and terahertz regimes. They found that the transition in relative orientations between the B-field and the filament elongation at about $A_\mathrm{V} = 21$ mag. Those authors claim that this configuration results from the B-field lines being dragged by the infalling material, feeding the central hub. To determine the relation between kinematics and the magnetic direction of the filament (especially the bending part of the B direction), further observations of \co with better sensitivity and spectral resolution are needed.

\begin{figure}
   \centering
   \includegraphics[width=\hsize]{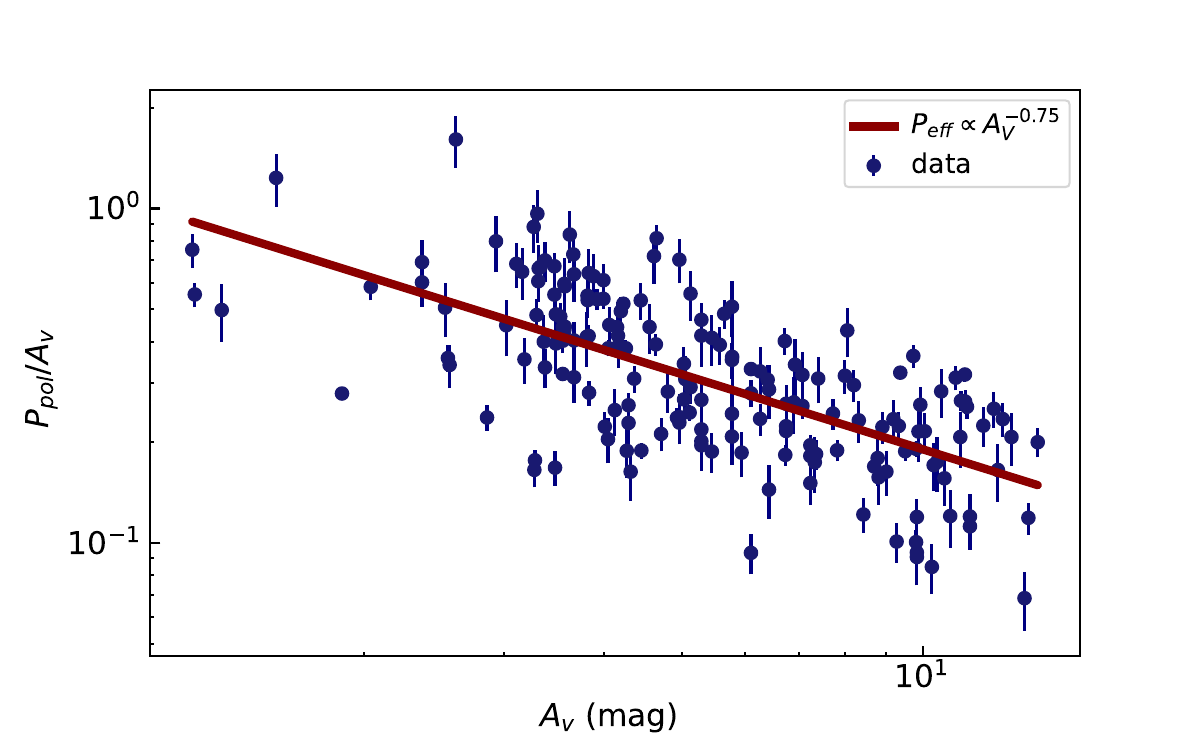}
      \caption{ Scatter plots of the polarization efficiency ($P_{\rm pol}/A_{\rm V}$) as a function of the visual extinction ($A_{\rm V}$) in magnitudes in logarithmic scale. Measurement uncertainties are displayed
as an error bar for each data point. The solid line is the best fit for the dataset, as explained in the main text. The best fit is shown in the top-right corner.}
         \label{efficiency}
   \end{figure}

Figure \ref{efficiency} shows the scatter plot of the polarization efficiency, which is described as the polarized fraction normalized by the visual extinction ($P_\mathrm{eff} = P_\mathrm{pol}/A_\mathrm{V}$), as a function of visual extinction ($A_\mathrm{V}$) in log-log space.
As a result of fitting a linear relationship between $P_\mathrm{pol}/A_\mathrm{V}$ and $A_\mathrm{V}$ (red curve in Fig. \ref{efficiency}), we obtain the slope of the relation $P_\mathrm{eff} \propto A_\mathrm{V}^\mathrm{-\alpha}$ (see Fig. \ref{efficiency}; the error bars on each data points are calculated from the least square solution). 
For our NIR data, we find a slope of $\alpha = 0.75 \pm 0.07$, which is consistent with radiative torque alignment (\citealt{lazarian2007}) theory and previous studies. According to \cite{alves14}, for NIR polarization, the values of $\alpha$ in the starless object Pipe-109 in the Pipe Nebula are $1$ and $0.34$ for $A_\mathrm{V} < 9.5$ mag and $A_\mathrm{V} > 9.5$ mag, respectively.
 \cite{redaelli19} calculated a very steep slope, $\alpha$ = 1.21, for far-infrared (FIR) polarization data on the protostellar core IRAS 15398-3359. 
 These slopes indicate depolarization at high column densities. Consequently, if dust grains are aligned by radiative torque originating from the interstellar radiation field at higher visual extinction, less radiation penetrates the cloud, and, therefore, grain alignment is less efficient. In addition, the dust grain size distribution has a great impact on alignment efficiency (\citealt{brauer16}; \citealt{pelko09}).

\begin{figure}
    \centering
    \subfloat{{\includegraphics[width=8.5cm]{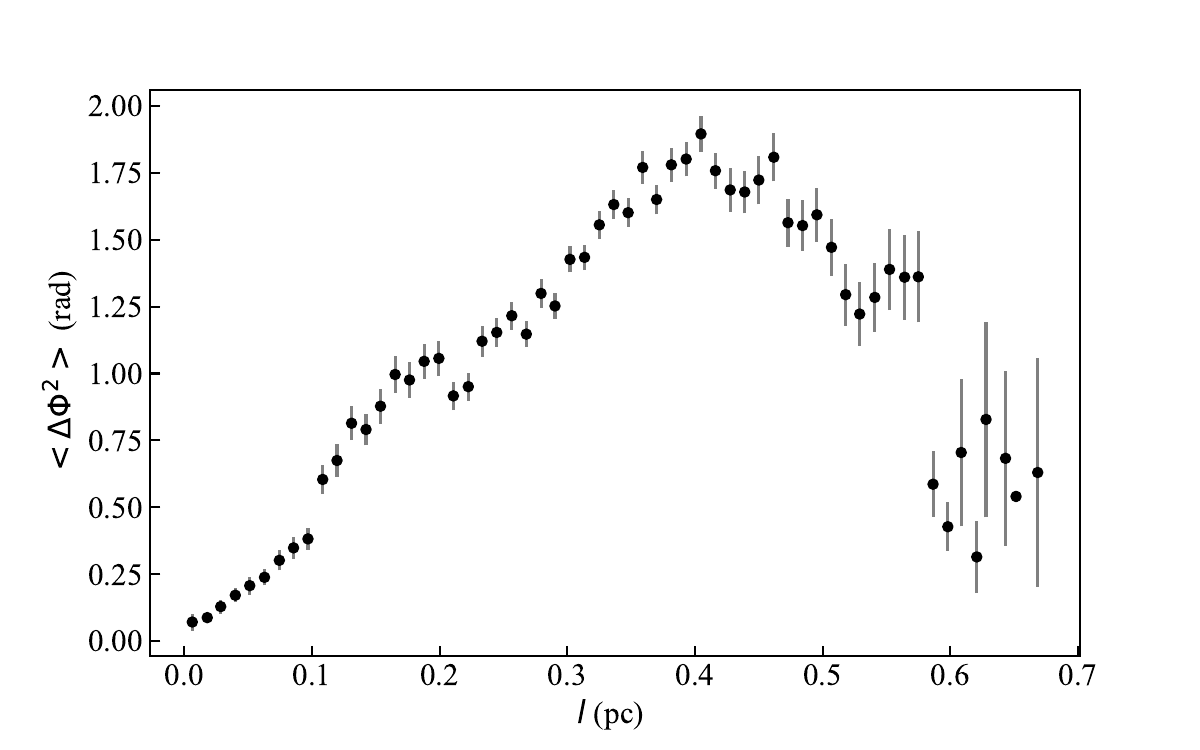} }}
    \qquad
    \subfloat{{\includegraphics[width=8.5cm]{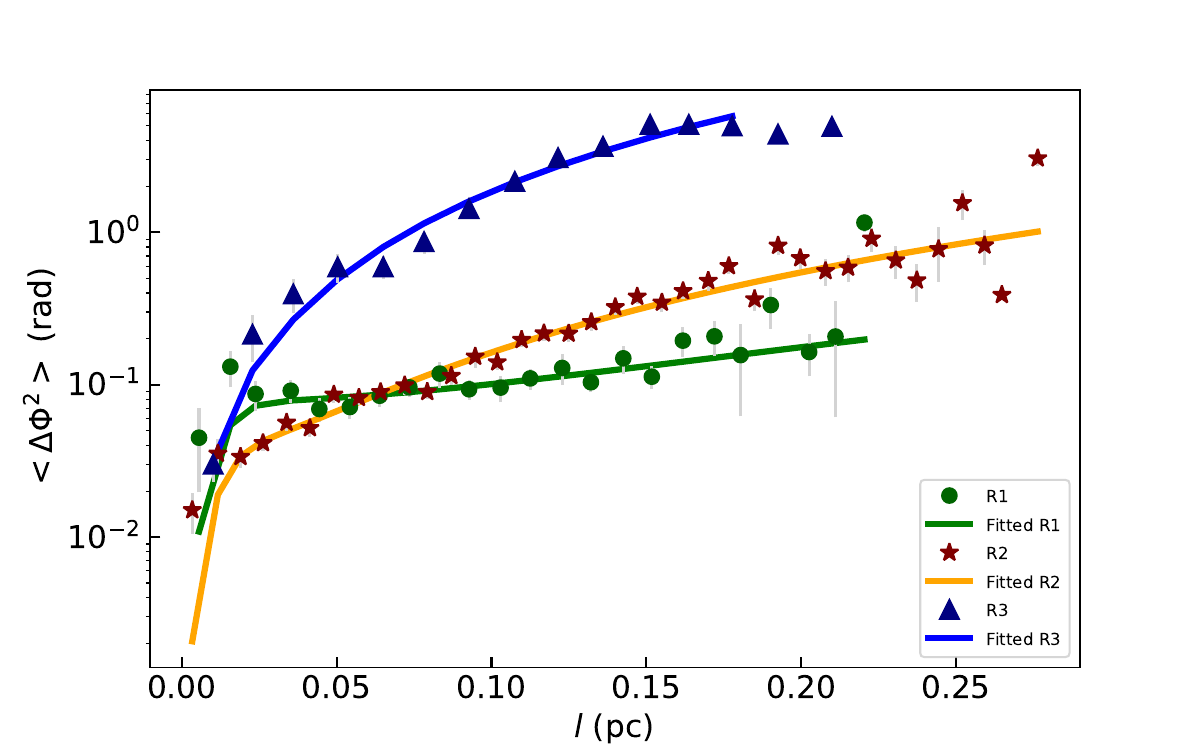} }}
    \caption{ ADF versus distance, $l$. Top panel: ADF for the whole filament. Bottom panel: Same but for the three filament regions. The best fits to the data points using Eq. (\ref{Houde}) for R1, R2, and R3 are shown with the green, orange, and blue curves, respectively. Measurement uncertainties are displayed as error bars, which are not visible for small values of $l$ due to the higher statistics. }
    \label{disper}
\end{figure}

\subsection{Dispersion of polarization angles and the magnetic field strength }
\label{sec2}

Various methods have been proposed to estimate the B-field strength from dust polarization data. According to \cite{davis} and \cite{chan-ferm}, hereafter DCF, the kinetic energy of turbulent motions is equal to the fluctuating magnetic energy density: 
\begin{equation}
\label{condition}
\frac{1}{2} \, \rho \, \delta \varv^2 = \frac{\delta B^2}{8 \pi}.
\end{equation}

The mean B-field ($B_0$) is
\begin{equation}
\label{B}
 B_0 = \sqrt{\frac{4\pi \rho}{3}} \frac{\delta \varv}{\delta \theta},
\end{equation}
where $\rho$  is the gas density, $\delta \varv $ is the velocity dispersion, and $\delta \theta$ is the dispersion of the polarization angle equal to the ratio of the turbulent B-field over the mean B-field ($\delta \theta = \delta B / B_0$).
\cite{hildebrand} and \cite{houde} developed the angular dispersion function (ADF) approach as an alternative way to estimate the $ \delta B/B_0 $ ratio (known as the HH09 method). 
The method was formulated to eliminate inaccurate estimates of the B-field generated by factors other than MHD waves, such as large-scale bending caused by differential rotation and gravity.
We calculated the autocorrelation function of the position angles, $\Delta \Phi$. This refers to the variation in angle between every pair of vectors separated by a distance of $l$ (see \citealt{houde}),

\begin{equation}
\label{position_angle}
 \langle \Delta \Phi^2(l)\rangle = \frac{1}{N} \sum_{i=1}^N [\Phi(x) - \Phi(x+l)]^2,
\end{equation}
which results in the following analytical relation,
\begin{equation}
\begin{split}
\label{Houde}
\langle \Delta \Phi^2(l)\rangle = 2 \sqrt{2 \pi} & \left( \frac{\delta B}{B_0}\right)^2  \frac{\delta^3}{(\delta^2 + 2W^2)\Delta^\prime}  \\
 & \times \left[ 1-\mathrm{exp} \left( -\frac{l^2}{2(\delta^2+2W^2)}\right) \right] + m^2 l^2,
\end{split}
\end{equation}
where $m$ is a constant related to the large-scale structure of the B-field and is not influenced by turbulence, $\Delta^\prime$ is the effective cloud depth, and $W$ is the telescope beam size. This relationship only applies to spatial scales $\delta \le l \le d$, where $\delta$ represents the correlation length of the turbulent B-field ($\delta B$), and $d$ represents the highest distance at which $B_0$ remains uniform.
The effective angular resolution of absorption dust polarization data is treated as zero (\citealt{franco10}). As a result, the angular resolution term in the equation is $W=0$. 
To obtain $ \delta B / B_0$ by fitting Eq. \eqref{Houde}, we calculated $ \Delta \Phi^2(l)$ for all available pairs of points. The spatial scale, $l$, is divided into 24, 38, and 16 bins corresponding to $\sim$ 0.009 pc, 0.007 pc, and 0.013 pc for the R1, R2, and R3 regions, respectively.
The values of the parameters $\Delta^\prime = 0.1 $ pc and $\delta = 10 $ mpc that were used in Eq. \eqref{Houde} are obtained from \cite{franco10}. They employed $\Delta^\prime = 0.3 $ pc for the whole region of B59, but since we only focused on the filament area, and each part of the filament has the length of $\sim$ 0.1-0.3 pc, we used  $\Delta^\prime = 0.1 $ pc. A $\Delta^\prime = 0.3 $ pc will decrease the magnetic strengths on the plane of the sky (PoS) ($B_\mathrm{pos}$) of each region by $\sim$ 24\% ( $B_\mathrm{pos}$ values in Table \ref{Parameters-1}).

By employing the $\textsc{Scipy}$ package, specifically using the $\textsc{Scipy.optimize.curve\_fit}$ module with least squares optimization, we found the best-fit solution of Eq. \eqref{Houde} in the three regions along with their associated errors, by leveraging the standard deviation of the fitted parameters obtained from the covariance matrix returned by the  $\textsc{curve\_fit}$ function. 
We estimated the strength of the B-field in each region (see Table \ref{Parameters-1}) using a further expansion of Eq. \eqref{B} (which is valid if $\delta \phi <25 ^\circ$; \citealt{crutcher}):
\begin{equation}
\label{strenght}
B_\mathrm{pos} = \sqrt{4 \pi \: \mu_\mathrm{H_2} \: m_\mathrm{H} \: n_\mathrm{H_2}} \frac{\delta v}{\delta \theta},
\end{equation}
where $m_\mathrm{H}$ is the hydrogen mass, $\mu_\mathrm{H_2} $ is the gas mean molecular weight per hydrogen molecule, and $n_\mathrm{H_2}$ is the gas volume density. We employed the $n_\mathrm{H_2} = 3 \times 10^3 \, \mathrm{cm}^{-3}$ from \cite{alves08}. They evaluated the gas volume density associated with the optical polarization in the different parts of the Pipe Nebula region.
We computed the gas velocity dispersion by fitting one Gaussian profile to the \co(3-2) molecular line, obtaining the mean value of $\delta \varv = 0.14 \pm 0.02 $ \kms in the filament region (see Sect. \ref{sec-kin}).

We calculated these parameters for all pairs of data points, obtaining the results shown in Fig. \ref{disper} (top panel). Our scatter plot does not show the expected increase in the ADF with distance $l$. In fact, due to the sampling of the data, we see peaks in the profile of the ADF.
The entire field of view was sampled at 0.08 degrees per tile, equivalent to $ 0.2$ pc, which corresponds to a peak every $ \sim 0.2 $ pc in Fig. \ref{disper}, top panel (see \citealt{soler2016} for more details on this effect). 
To reduce these spurious effects, we did our calculation for each region separately, obtaining the fit parameters for every region of the filament (see Fig.\ref{disper}, bottom panel).
It is evident that the distribution of $\Delta \Phi^2$  along the filament follows a clear trend. In Fig. \ref{disper} the R3 region has a steeper slope compared to other regions. A steep slope is evidence that the B-field orientation in the plane of the sky changes significantly. As the slope increases from R1 to R3, the dispersion of position angles becomes larger.
Table \ref{Parameters-1} shows the best-fit parameters for Eq. \eqref{Houde} and the plane of the sky magnitudes of the B-field for each region. The B-field strength increases from R1 to R3 (see the fourth column of Table \ref{Parameters-1}), ranging from 151 to 234 $\mu$G.

 
\begin{table*}[ht]

\caption{ Physical parameters for the three regions of the filament from Eq. \eqref{Houde}.} 
\centering 
\renewcommand{\arraystretch}{1.1}

\begin{tabular}{c c c c c c c} 
\hline \hline
\toprule

Region & $ \delta B / B_0$ & $m$ [rad/pc] & $B_\mathrm{pos}$ [$\mu$G]& $B_\mathrm{pos}^{*(a)}$ [$\mu$G] 
 & $N(\mathrm{H}_2) [\mathrm{cm}^{-2}]$ &$\lambda ^{(b)}$  \\  
\midrule

R1 & $0.39\pm 0.02$  & $ 1.6\pm 0.2$  & $151 \pm 6$ & $66 \pm 1$ & 4.1 $\times 10^{21}$ & $0.27 \pm 0.04$ \\    
R2 & $0.26\pm 0.01$  &  $3.57\pm 0.06$  & $222\pm 9$ & $80 \pm 1$  & 3.9 $\times 10^{21}$ & $0.13 \pm 0.02$ \\
R3 & $0.25\pm 0.06$  &  $13.5 \pm 0.2$ & $234 \pm 6 $ & $83 \pm 8$  &3.9 $\times 10^{21}$ & $0.13 \pm 0.02$  \\  
   
\hline \\ 
\end{tabular}
\label{Parameters-1} 
\\
\footnotesize \textbf{Note}: $^{(a)}$ $B_\mathrm{pos}^{*}$ is the B-field strength calculated from \cite{skalidis21}. $^{(b)}$  $\lambda$ is cloud stability; see Sect. \ref{stab}.  In all computations, we considered $\delta = 10$ mpc, obtained from \cite{franco10}.  \\

\end{table*}

\subsection{ The influence of assumed parameters and used methods}

The previous analysis relied on specific key parameters. In this section we examine if altering these assumptions can influence our findings. Specifically, the initial focus is be on the turbulence correlation scale, $\delta$. For the calculation in Sect. \ref{sec2}, we used $\delta = 10 $ mpc, considering that  \cite{franco10} reported a turbulence correlation scale of a few milliparsecs for the Pipe nebula. The values found align well with previous estimates of this parameter in other areas where star formation occurs. For example, \cite{coud19} measured a value of $\delta$ = 7 mpc in the Barnard 1 region. In contrast, in the high-mass star-forming region NGC 7538, \cite{frau14} calculated a range of values for $\delta$ between 13 and 33 mpc, and \cite{redaelli19} investigated four values of 5, 10, 15, and 20 mpc for IRAS 15398-3359, showing an increase in the B-field by increasing $\delta$. 
We therefore used Eq. \eqref{Houde} again, but this time obtaining the best solutions for three free parameters, $m$, $ \delta B / B_0$, and $\delta$ (the same method as explained in Sect. \ref{sec2}). The B-field strength is calculated for each region based on these updated parameters. Table \ref{Parameters2} shows the best-fit parameters and B-field strengths. We obtain the same turbulence correlation value ($\delta = 10$ mpc) and B-field strength ($B = 151$ $\mu$G) for the R1 region, while for the R2 region is lower than before (with two free parameters; see Table \ref{Parameters-1}), and for R3 region the fit routine does not converge, due to the limited number of independent data points available.

\begin{table}[ht]
\caption{ Best-fit parameters with three free parameters, using Eq. \ref{Houde}. } 
\centering 
\renewcommand{\arraystretch}{1.1}

\begin{tabular}{c c c c c } 
\hline \hline
\toprule
Region & $ \delta B / B_0$ & $m$ & $\delta$  & $B_\mathrm{pos}$    \\  
&  &  (rad/pc) &  (mpc) & ($\mu$G) \\[0.7ex]

\midrule
R1 & $0.4\pm 0.1$  &  $1.6\pm 0.2$  & $10\pm 4$& $151\pm 9$ \\

R2 & $0.46\pm 0.05$  &  $3.6\pm 0.1$  & $3.0\pm 0.01$& $127\pm 4$ \\

   
\hline \\ 
\end{tabular}
\label{Parameters2}

\end{table}

Next, we discuss the impact of assumption on the methods we used to calculate the $B$ strength.
In this work, we used the combination of DCF and HH09 methods to calculate the B-field strength of the filament. Various modifications to the DCF method have been proposed (e.g., \citealt{ostriker01}; \citealt{houde}; \citealt{cho-yoo}; \citealt{lazarin20}; \citealt{skalidis21}).
There are several assumptions made by DCF, including the fact that turbulence is Alfv\'enic, as well as that $\sigma_{\rm nt}$ traces turbulent motions and that the mean B-field is uniform. In general, it has been suggested that the original DCF method overestimates B-field strength (\citealt{pattle22}; \citealt{ostriker01}; \citealt{houde}; see \citealt{pattle19} and \citealt{skalidis21} for a thorough comparison). To show the differences, we used one of the modification methods suggested by \cite{skalidis21}. 
Their modified equation includes non-Alfv\'enic (compressible) modes and Alfv\'enic (incompressible) modes and leads to
\begin{equation}
\label{skalidis}
B_\mathrm{pos}^* = \sqrt{2  \, \pi \, \rho} \, \frac{\delta \varv}{ \sqrt{\delta \theta}}.
\end{equation}
Those authors claim that this method is more accurate in retrieving $B_\mathrm{pos}$, and has lower error than DCF and HH09 methods. Table \ref{Parameters-1} shows calculated values for each region. As expected, the $B_\mathrm{pos}$ for all regions has values lower than those obtained with the DCF and HH09 methods.

\subsection{Cloud stability }
\label{stab}
Determining the importance of B-fields for cloud gas dynamics and cloud balance is one of the main motivations for studying the B-field in the cloud. However, determining how the B-field affects a cloud's dynamic behavior remains challenging. One parameter that helps us quantify this importance is the mass-to-flux ratio normalized to the critical value, $\lambda$. Observed mass-to-flux ratios exceeding the critical value of $1$ indicate that the cloud is supercritical, and the B-field cannot prevent it from collapsing. On the other hand, clouds are subcritical if their ratio is $\lambda<1$ and the B-field stabilizes gravity. We calculated $\lambda$ from (\citealp{crutcher})

\begin{equation}
\label{landa}
\lambda  = \frac{(M/\Phi)_\mathrm{observed}}{(M/\Phi)_\mathrm{crit}} = 7.6 \times 10^{-21} \frac{N(\mathrm{H}_2)}{B}   \left(\frac{\mu G}{\mathrm{cm}^{-2}}\right),
\end{equation}
where $N(\mathrm{H}_2)$ is the column density of the filament in units of \cm, and $B$ is the total B-field strength given in $\mu$G.
We used $B = B_\mathrm{pos}$ derived from Eq.\ref{strenght}. We note that this introduced a source of uncertainty, as the inclination of the total B field with respect to the PoS.
We used the $N(\mathrm{H}_2) $ values present in Table \ref{Parameters-1}, obtained from the \textit{Herschel} column density map, which represents the average $N(\mathrm{H}_2)$ for each region.
 In the last column of Table \ref{Parameters-1}, we report our derived values of $\lambda$ for each region. The results for all three regions in the filament indicate that their inferred mass-to-magnetic flux ratios are subcritical (see Table \ref{Parameters-1}).It is worth emphasizing that, despite employing $B_\mathrm{pos}^*$ <$B_\mathrm{pos}$), the parameter $\lambda$ consistently remains below 1. This means that B-fields are dynamically relevant and are expected to support the filament against its own gravity, especially for the region R3. The B-field in R3 is probably distorted by the gravity of the nearby central hub. 
However, one should keep in mind that these considerations do not take into account corrections for the geometry of the filament and the B-field. Strictly speaking, the mass $M$ in Eq. \ref{landa} should be that contained in a flux tube characterized by a B-flux $\Phi$. The derived values of the mass-to-flux ratio should therefore be taken as indicative.


\section{ Modeling of the filament}
\label{sec:model}

%

We used the model developed by TCa and TCb in order to interpret the properties of the filament of B59. 
TCb considers the stability of a self-gravitating, polytropic cylinder threaded by a helical B-field as a function of the relative strength of the axial and azimuthal components of the field (see also \citealt{fiege2000}). Although our polarization maps suggest the presence of a twist in the B-field (see Fig. \ref{kinematics} or Fig. \ref{extinction}), our data are insufficient for a detailed comparison with the models of TCb, and we defer this analysis to subsequent work. 
Here, we present a preliminary analysis following TCa, where the extra support provided by B-fields (and possibly turbulence) against gravity is modeled by assuming a polytropic relation between the gas pressure, $p,$ and the gas density, $\rho$,

\begin{equation}
p = K  \:  \rho^\mathrm{\gamma_\mathrm{p}},
\end{equation}
where $\gamma_\mathrm{p}$ is the polytropic exponent and $K$ is a constant.

\begin{figure}
    \centering
    \includegraphics[width=1\linewidth]{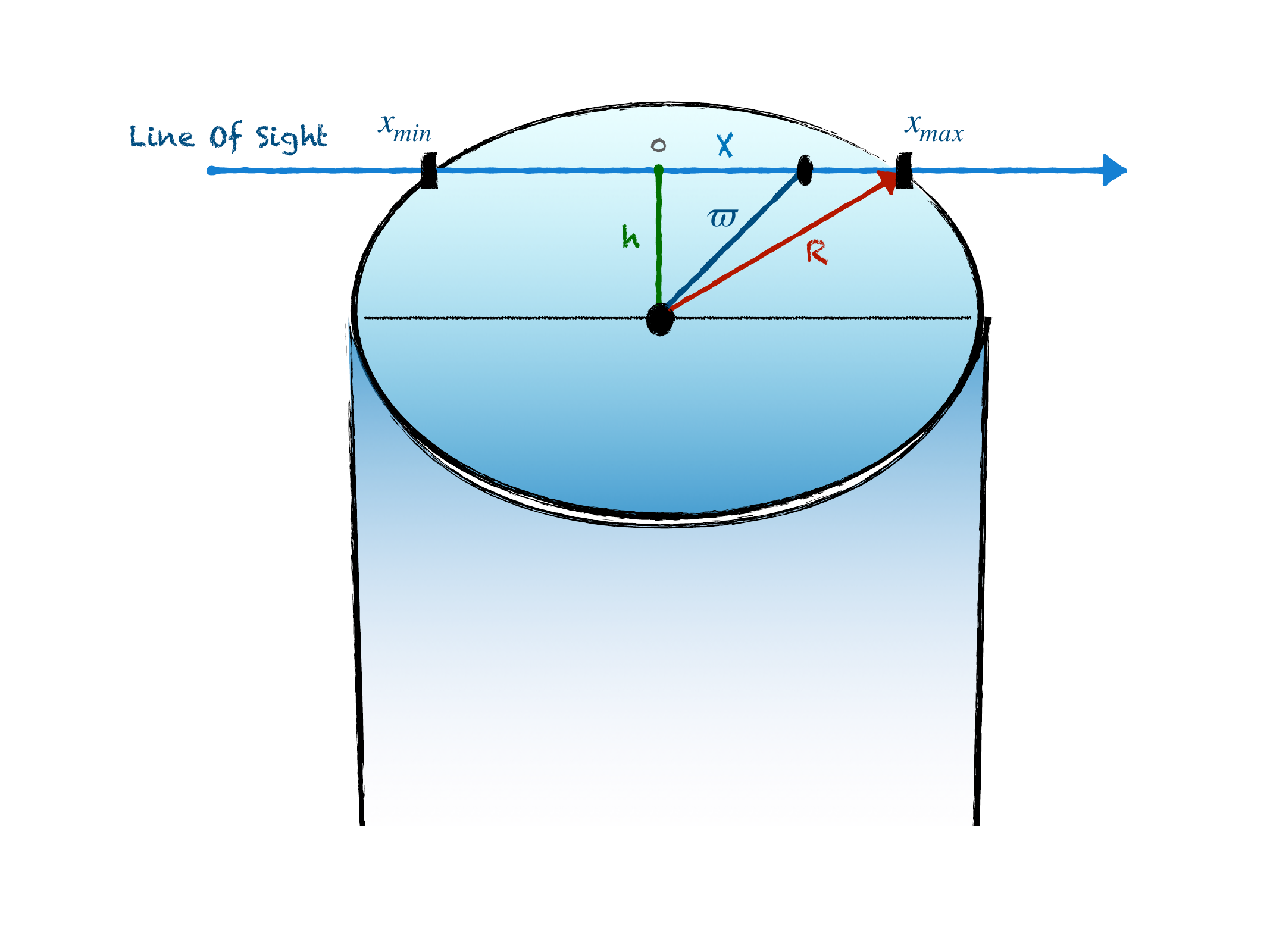}
    \caption{Simplified model of a cylindrical filament illustrating all parameters used for modeling. The cylinder is assumed to have an infinite length and radius $R$ (see Eq. \ref{eq-non-diem} and Eq. \ref{eq-density}). The blue arrow indicates the direction of the observer's line of sight. }
    \label{cartoon}
\end{figure}

\begin{figure}
    \centering
    \subfloat{{\includegraphics[width=8.5cm]{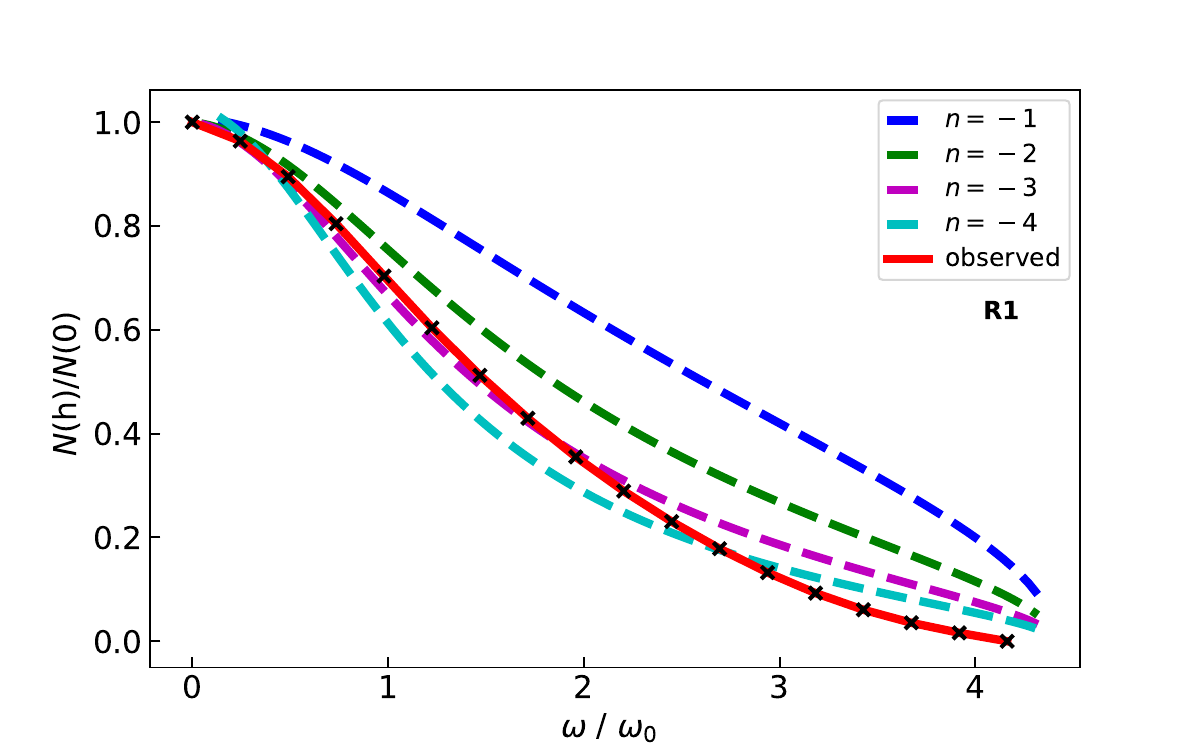} }}
    \qquad
    \subfloat{{\includegraphics[width=8.5cm]{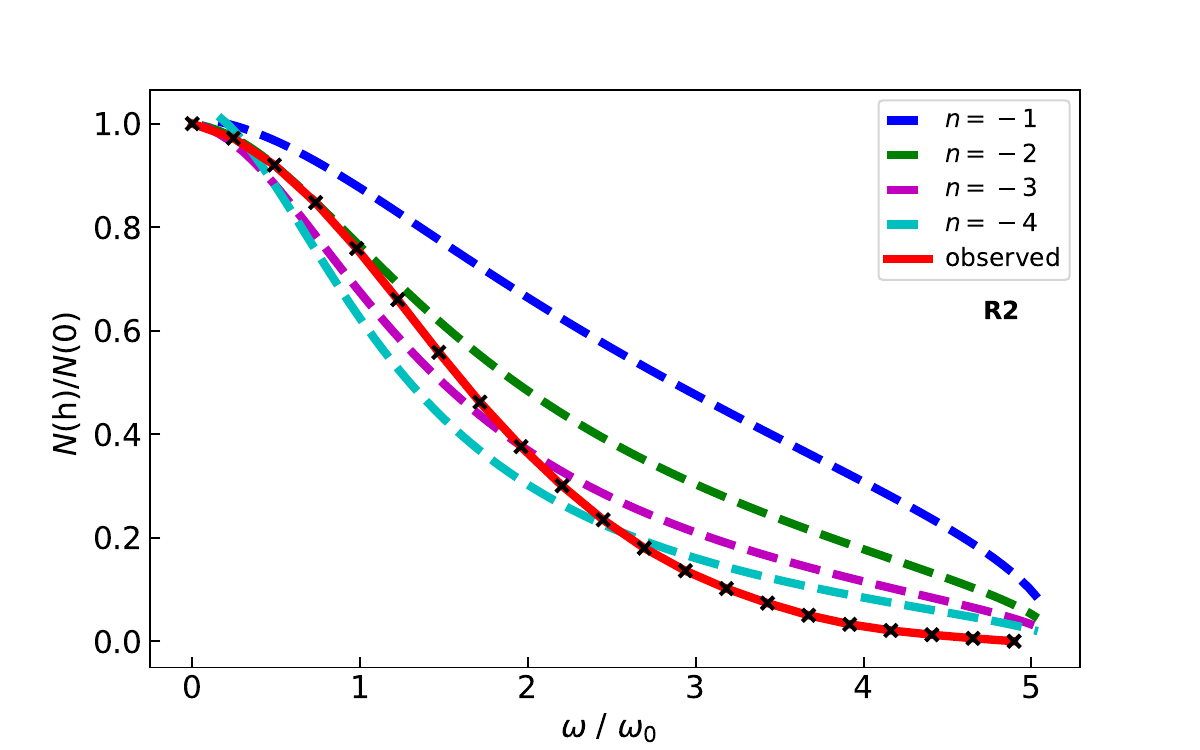} }}
    \caption{ Radial column density profiles (with background subtraction and normalized to the central column density, $N(0)$) of both observations (solid red curves) and polytropic cylinder models with $n = -1, -2, -3, -4$ (dashed curves). The radius is normalized to $\varpi_0$. The black crosses indicate the observation data points. The top and bottom panels describe regions R1 and R2 of the filament, respectively.}
    \label{model_1}
\end{figure}

In this case, we assumed that the filament is perpendicular to the line of sight and is cylindrically symmetric. 
In a polytropic cylinder, the temperature $T_{\rm p} \propto (dp/d\rho)^{1/2}$ (hereafter the polytropic temperature) is a measure of the radial support against gravity. The comparison of the measured gas temperature $T_{\rm g}$ with $T_{\rm p}$ therefore reveals whether or not the thermal pressure gradient is sufficient to provide support to the cloud.
In all calculations, we used polytropic index $n$, defined as $ \gamma_\mathrm{p} = 1+ 1/n$. We implemented the TCa models to describe the filament in B59.
TCa speculated about the support given by a superposition of low-amplitude Alfv\'en waves, which behave as a polytropic gas with $n=-2$, resulting in density profiles in agreement with observations. Fully developed MHD turbulence is another possibility, although these filaments are not very turbulent, as shown in Sect. \ref{sec-kin}. 
Considering the curved shape of the filament, we divided the filament into three regions (R1, R2, and R3; see Fig. \ref{extinction}), as explained in Sect. \ref{sec-pola}. 
In the R1 region, most of the polarization orientations are generally perpendicular to the cylinder spine, while in the R2 region, the orientation tends to align with the cylinder spine. Lastly, the R3 region has a more complex polarization pattern resembling a U-shape.
In each segment, the observed column density $N({\rm H}_2)$ was evaluated as a function of cylindrical radius, averaging the two sides from the center to the edges (0 to $x_\mathrm{max}$ and 0 to $x_\mathrm{min}$; see Fig. \ref{cartoon}). 
Figure \ref{model_1} shows the results for the R1 and R2 regions of the filament.

The column density profile obtained in this way from $\emph{Herschel}$ data was compared to the column density profiles of unmagnetized polytropic filaments computed by TCa for various cylindrical polytropes index (see Fig. \ref{model_1}).
To compare the observation data with the model from TCa, we converted the density of the model to column density $N(h)$.
As a means of making the calculation more straightforward, we used the standard nondimensional density, $\theta$, and radius, $\xi$, parameters defined as
\begin{equation}
\label{eq-non-diem}
\varpi = \varpi_0 \xi  ,\: \:  \;  \rho = \rho_0 \theta^n 
,\end{equation}
where $ \varpi_0 =   \sqrt{ c_s^2 / (4 \pi G \rho_0)}  $ and $\rho_0$ is the density at the center of the filament (for more details about the model, see TCa). We assumed volume number density on the axis $n_c = 2 \times 10^4 $ cm$ ^{-3}$   to calculate $\rho_0= \mu \, m\mathrm{_H} \, n\mathrm{_c}$.
In the models, the column density,$N(h)$, was computed from the tabulated density profile $n(\varpi)$, as

\begin{equation}
\label{eq-density}
N(h)= \int_h^R \frac{n(\varpi)\varpi}{\sqrt{\varpi^2-h^2}}\, d\varpi = \frac{2 \: \rho_0 \: \omega_0}{\mu \: m_\mathrm{H}} \int_{h/\varpi_0}^{R/\varpi_0} 
\frac{\theta^\mathrm{n} \: \xi }{\sqrt{\xi^2-h^2/\varpi_0^2}}\, d\xi,
\end{equation} 
where $R$ is the radius of the filament's ``surface,'' where the filament merges into the ambient medium (also called the ``truncation radius''; see Fig. \ref{cartoon} for more details).
A radial column density profile of observed and polytropic cylinder models for various $n$ is shown in Fig. \ref{model_1}. Column densities are normalized to on-the-axis values, $N(0)$, to eliminate biases due to arbitrary choice of $n_c$. 
The averaged column density profiles of the R1 and R2 regions are well reproduced by cylindrical polytropes with $-4 \lesssim n \lesssim -2$. In general, a single value of $n = -3 $ is able to provide a good fit to the data in regions R1 and R2. We used $n = -3$ as the best-fit value in the rest of the analysis.
The fitted parameters for these regions are listed in Table \ref{Parameters-4}. In R3, the fit is not good because it is difficult to consider this part of the filament as cylindrically symmetric. Furthermore, the R3 region lies close to the central hub, and is possibly characterized by strong gas flows, which makes it more challenging to fit with our simplified model.


According to the set of parameters for R1 and R2 regions (see Table \ref{Parameters-4}), the stability parameter $\xi_\mathrm{s} = R/\varpi_0  = 4.16$ and 3.9 is smaller than the critical value $\xi_\mathrm{cr} = 4.28$ for R1 and R2 regions, respectively (see Table 1 of TCa). Given that $\xi_s$ and $\xi_{cr}$ are very close, and considering the various assumptions made in the model, we conclude that $\xi_s$ is not significantly larger than $\xi_{cr}$, and there is no sign of instability in the filament. Therefore, the filament is stable to radial collapse.

\begin{table*}[ht]
\caption{ Model parameters.  }
\centering 
\renewcommand{\arraystretch}{1.1}

\begin{tabular}{c c c c c c c c c c c} 
\hline \hline
\toprule

Regions & $n$ & $\gamma_p$ & $\varpi_0$ (cm) & $\varpi$ = $R$ (cm) & $\xi_{cr} ^{(a)}$ & $\xi_s = R$/$\varpi_0$  \\  

\midrule

R1 & $-3$  & 2/3   & 7.25$\times 10^{16}$  & 3.02$\times 10^{17}$  & 4.28 & 4.16\\    

R2 & $-3$  &  2/3  &  7.25$\times 10^{16}$ & 2.84$\times 10^{17}$ & 4.28 & 3.92\\

\hline \\ 
\end{tabular}
\label{Parameters-4} 
\\
\footnotesize \textbf{Note}: $^a$ from \cite{toci1}.  \\

\end{table*}



Cylindrical polytropes have a different mass per unit length depending on the polytropic index. We calculated the mass per unit length for various polytropic indexes using the following integral:

\begin{equation}
\frac{M}{L} = 2 \pi \rho_0 \varpi_0^2 \int_0^{R/\varpi_0} \theta^n \xi d\xi.
\label{ml}
\end{equation} 
The estimated $\frac{M}{L}$ for each model are listed in Table \ref{Parameters3}.
In Sect. \ref{sec:mass} we show that the $\left(\frac{M}{L}\right)_\mathrm{fil}$ from the observation data is smaller than the critical value, which indicates the filament is in a stable state. 
We show in Table \ref{Parameters3} that $\frac{M}{L}$ (from Eq. \ref{ml}) for $n= -3$ is smaller than the critical value in agreement with observations.

\begin{figure}
    \centering
    \subfloat{{\includegraphics[width=8.5cm]{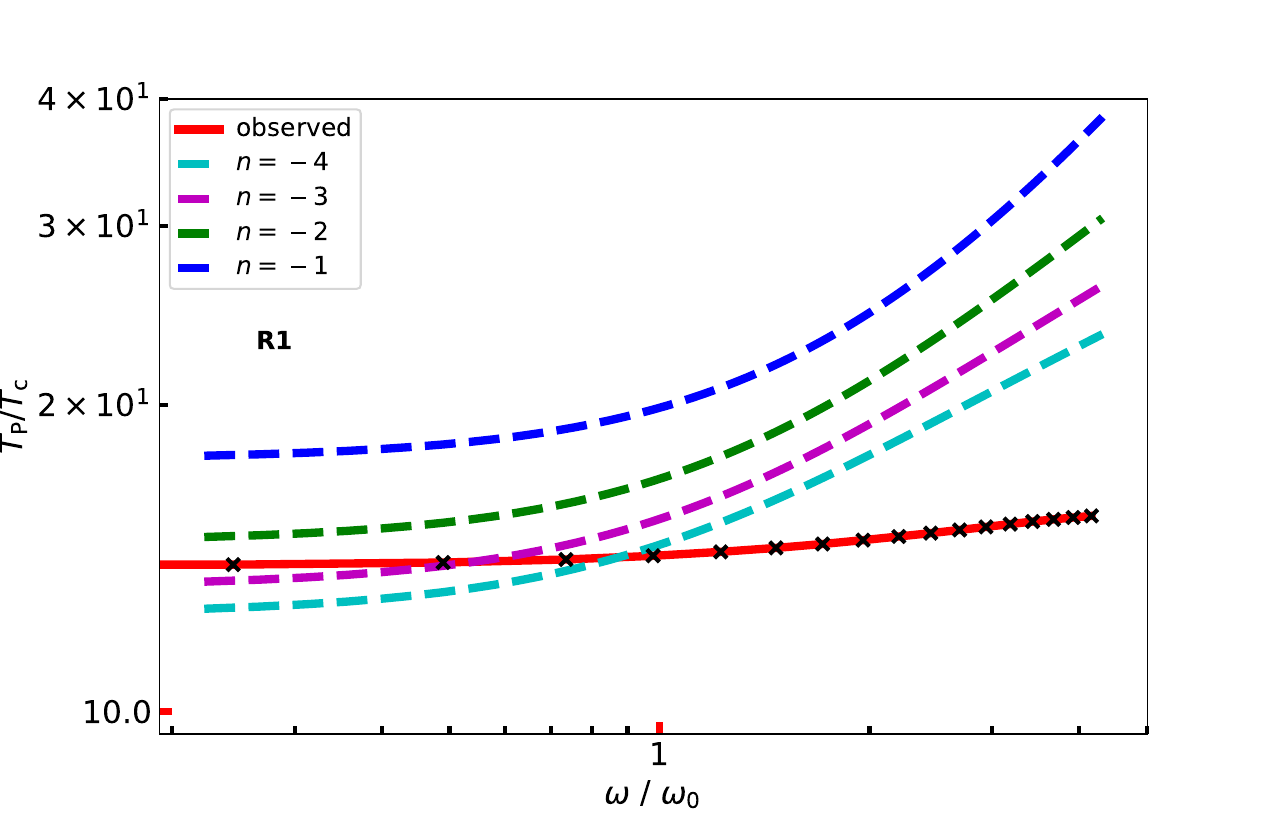} }}
    \qquad
    \subfloat{{\includegraphics[width=8.5cm]{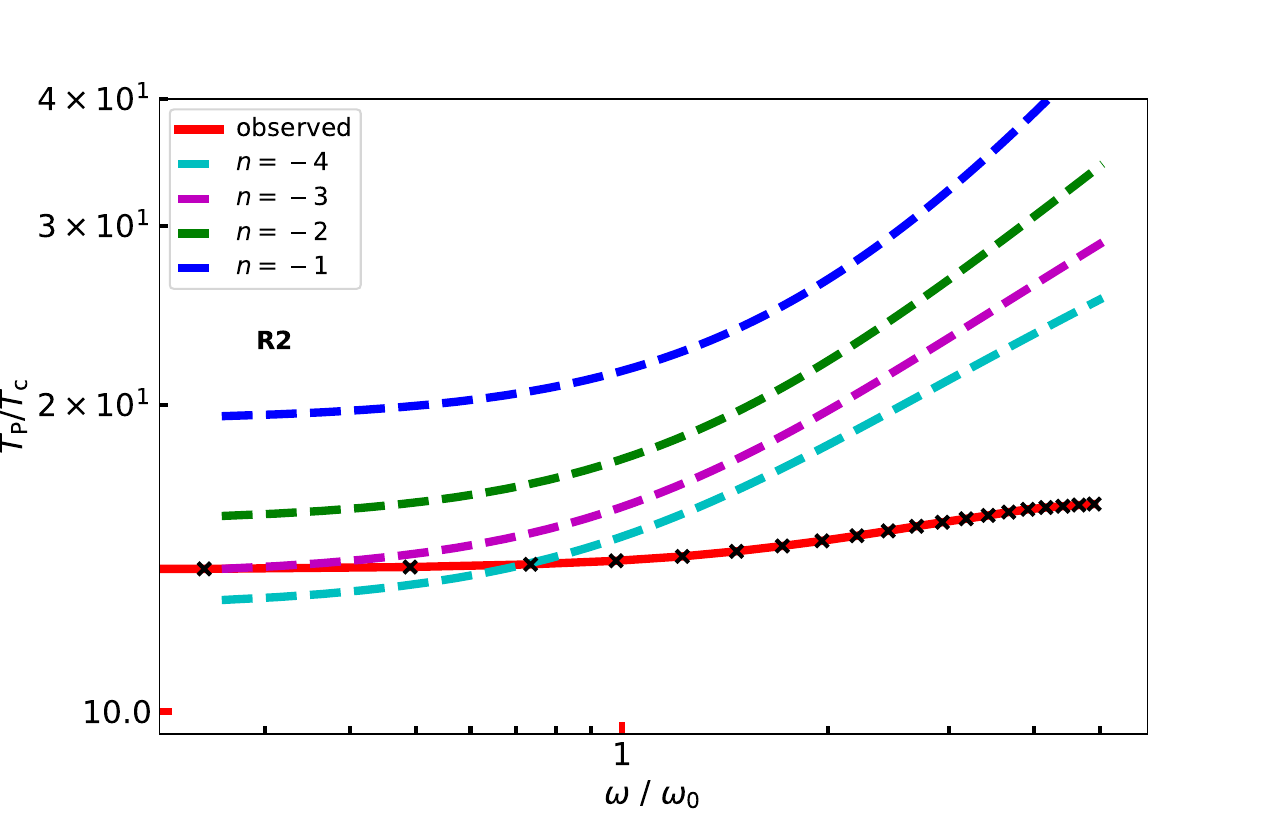} }}
    \caption{ Radial profiles of the observed temperature and the polytropic temperature, $T_{\rm P}$ (normalized to the central temperature value, $T_{\rm c}$, for the same polytropic indexes, $n$, as in Fig. \ref{model_1}). Solid and dashed lines represent observations and models, respectively. The black crosses indicate the observation data points. The top and bottom panels represent regions R1 and R2 of the filament, respectively.}
    \label{model_2}
\end{figure}

We used the data of Fig. 2 from the TCa model to plot radial profiles of the polytropic temperature $T_\mathrm{p}$ of a polytropic cylinder. 
In Fig. \ref{model_2} the dashed lines show the temperature as a function of radius from the models. In contrast, the solid red curves show dust temperature as a function of radius for the filament, obtained from $\emph{Herschel}$ data for R1 and R2. 
The dust temperature $T_\mathrm{d}$ increases outward from $\sim$ 13.9 K on the axis to $\sim$ 15.5 K at $\varpi \sim$ 0.09 pc in the region R1 of the filament (see Table \ref{tem_para}; the dust temperature map is shown in Appendix \ref{app-b}). As for the gas temperature $T_\mathrm{g}$, similar (or even steeper) gradients are expected. Gas and dust are expected to be coupled at high densities above $ \sim 10^5 \,$ cm$^{-3}$  (\citealt{goldsmith2001}), whilst gas temperatures in the outer more diffuse regions are expected to be higher than the dust temperature (\citealt{keto-case}; \citealt{Gal-Wal}). 
Figure \ref{model_2} presents radial profiles of the observed dust and polytropic temperature $T_{\rm p}$ for the same polytropic indexes shown in Fig. \ref{model_1}.
The polytropic temperature (a measure of the polytropic pressure, which includes all contributions to the support of the filament, not just thermal pressure) at the surface of the filament is a factor of $\sim$ 2
larger than on the center of cylinder for regions R1 and R2 (see Table \ref{tem_para}). The measured dust temperature for the same regions of the filament, from the center to the surface, increases by 12\% and 14\%,
respectively (from $\emph{Herschel}$ dust map). The gas temperature is not known; however, it is unlikely to increase by a factor as large as 1.90 or 1.84.
Assuming $T_\mathrm{g}(0)=T_\mathrm{d}(0)=14$ K, for thermal pressure alone to support the R1 part of the filament, the gas temperature at the surface of the filament must be $T_\mathrm{g}(R)=25.3$ K, which is unrealistic (and the same reason for the R2, $T_\mathrm{g}(R)=24.3$ K, Table \ref{tem_para}). Even if the profile for the  $\emph{Herschel}$ temperatures is flatter than the real profile, due to the averaging of different temperatures along the line of sight (which tends to overestimate the temperatures in the densest and coldest regions), a sharper increase toward the outskirts should be observable. Therefore, the filament must be supported radially by some other agent(s) than thermal pressure, like B-fields or (albeit less likely, as previously discussed) turbulent support.

\begin{table}[ht]
\caption{ Temperatures at the center and surface of a cylinder from models and observations ($T_{\rm d}$).} 
\centering 
\renewcommand{\arraystretch}{1.1}

\begin{tabular}{c c c c c } 
\hline \hline
\toprule
 & R1$_\mathrm{obs}$ & R1$_\mathrm{mod}$ & R2$_\mathrm{obs}$  & R2$_\mathrm{mod}$   \\  

\midrule
$T_\mathrm{c}^{(a)}$  (K) & 13.9   &  13.3   & 13.8 & 13.2  \\

$T_\mathrm{R}^{(b)}$  (K) & 15.5   &  25.3   & 15.7  & 24.3  \\
$R$  (pc) & 3.9   &  3.9  & 4.2  & 4.2   \\  
   
\hline \\ 
\end{tabular}
\label{tem_para} 
\\
\footnotesize \textbf{Note}: $^{(a)}$ Temperature at the center of the cylinder, \\$^{(b)}$ Temperature at the surface of the cylinder.  \\

\end{table}

\section{Conclusions}
\label{sec-concl}

 We used \co and \cco (3-2) transition data from JCMT to reveal the kinematics toward one filament in B59, and NIR polarization observations to investigate the polarization properties of this isolated filament. We modeled the filament as a self-gravitating hydrostatic cylinder, as in TCa. Our results are as follows:

\begin{itemize}
    \item According to the nonthermal motion strength derived from the \co line, the filamentary cloud appears to be in a quiescent state where the nonthermal gas motions are smaller than or on the order of the sound speed of the gas ($\mathcal{M} = 0.54 < 1$). On the contrary, for the \cco emission, which traces the larger-scale gas, we obtain $\mathcal{M} = 1.0$, indicating transonic motions.    
    \item Based on the  $V_\mathrm{lsr}$ map of \co, we determine a gas flow toward the location of the hub with a velocity gradient of about $ \sim 0.69 \pm 0.01 $ km s$^{-1}$ pc$^{-1}$.
    
    \item The orientation of the polarization angles changes from perpendicular to the filament spine in the eastern region to parallel to it going toward the west. This could be caused by local velocity flows of matter infalling toward the hub, with the B-field being dragged by gravity along the filaments.
    \item The distribution of the polarization angles shows two peaks. One Gaussian profile corresponds to polarization angles in the R3 region, while the other Gaussian distribution suggests a broader distribution due to the merger of the R1 and R2 regions.
    \item Using the DCF and HH09 methods, we calculated the strength of the B-field on the plane of the sky for three different regions of the filament; they range from 150 to 230 $\mu$G (see Table \ref{Parameters-1}).
    \item Our analysis shows that the mass-to-flux ratio is lower than the critical value (i.e., the filament is subcritical), which means that the B-field is nominally sufficient to stabilize gravitational collapse due to its own gravity.
    \item The observed radial density profiles of filaments are, in general, well represented by polytropic indexes in the range $-3 < n < -3/2$ (see Fig. 1 in TCa). This rules out isothermal gas pressure as a supporting agent for $n= \pm \infty$. In principle, an outward increasing gas temperature can provide the required pressure gradient, but the observed dust temperature gradient is too shallow (see Fig. \ref{model_2}). It is unlikely that the gas temperature gradient is much steeper at these densities.  
    The observed radial profiles of the density and temperature profiles of the filamentary cloud, compared to theoretical models of polytropic self-gravitating cylinders, show that the filament is stable to radial collapse and is supported radially by agents other than thermal pressure.
    
\end{itemize}

Our findings suggest that the filament remains stable against radial collapse and is supported radially by factors beyond thermal pressure. The B-field is nominally sufficient to stabilize gravitational collapse due to its own gravity. The B-field we derived is adequately strong to counteract gravitational pull from the filament. Our result of a magnetically subcritical condition implies that the filament is stable and not prone to fragmentation.


\begin{acknowledgements}
      E. Redaelli acknowledges the support from the Minerva Fast Track Program of the Max Planck Society. G. Franco acknowledges the partial support from the Brazilian agencies CNPq and FAPEMIG. A. Duarte-Cabral acknowledges the support from a Royal Society University Research Fellowship (URF/R1/191609). 
      This research has used data from the $\emph{Herschel}$ Gould Belt survey (HGBS) project (http://gouldbelt-herschel.cea.fr). The HGBS is a $\emph{Herschel}$ Key Programme jointly carried out by SPIRE Specialist Astronomy Group 3 (SAG 3), scientists of several institutes in the PACS Consortium (CEA Saclay, INAF-IFSI Rome, and INAF-Arcetri, KU Leuven, MPIA Heidelberg), and scientists of the $\emph{Herschel}$ Science Center (HSC),\cite{andre10}. 
      The authors thank Juan Diego Soler for his support and discussion about magnetic field morphology and Claudia Toci for several interesting discussions of the modeling work used in this study.
\end{acknowledgements}

%
%

\bibliographystyle{aa} 
\bibliography{aanda} 

\begin{appendix} 
\section{Residual of the Gaussian fit }
\label{appen1}
\begin{figure}[h!]
    \centering
    \subfloat{{\includegraphics[width=8.5cm]{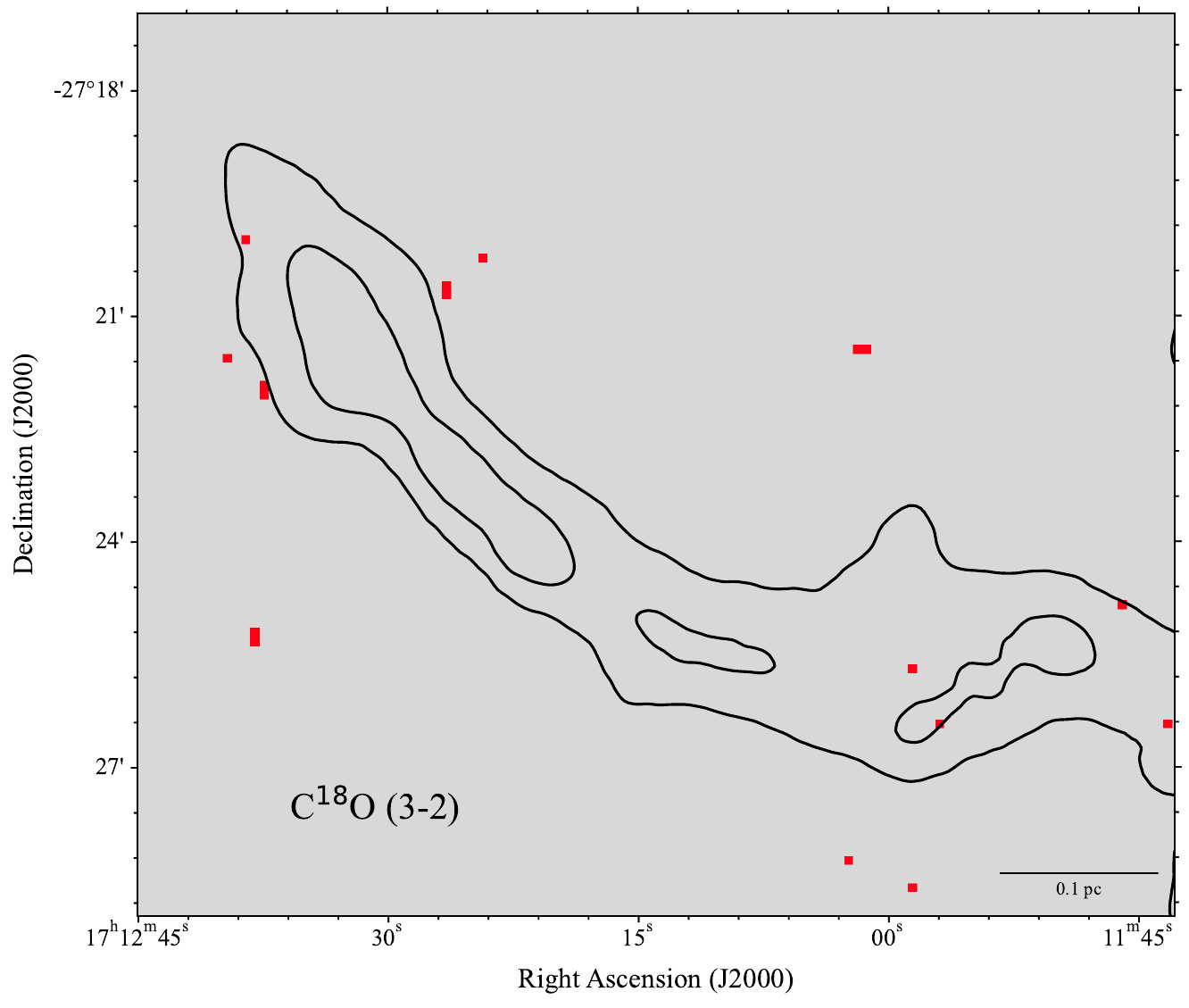} }}
    \qquad
    \subfloat{{\includegraphics[width=8.5cm]{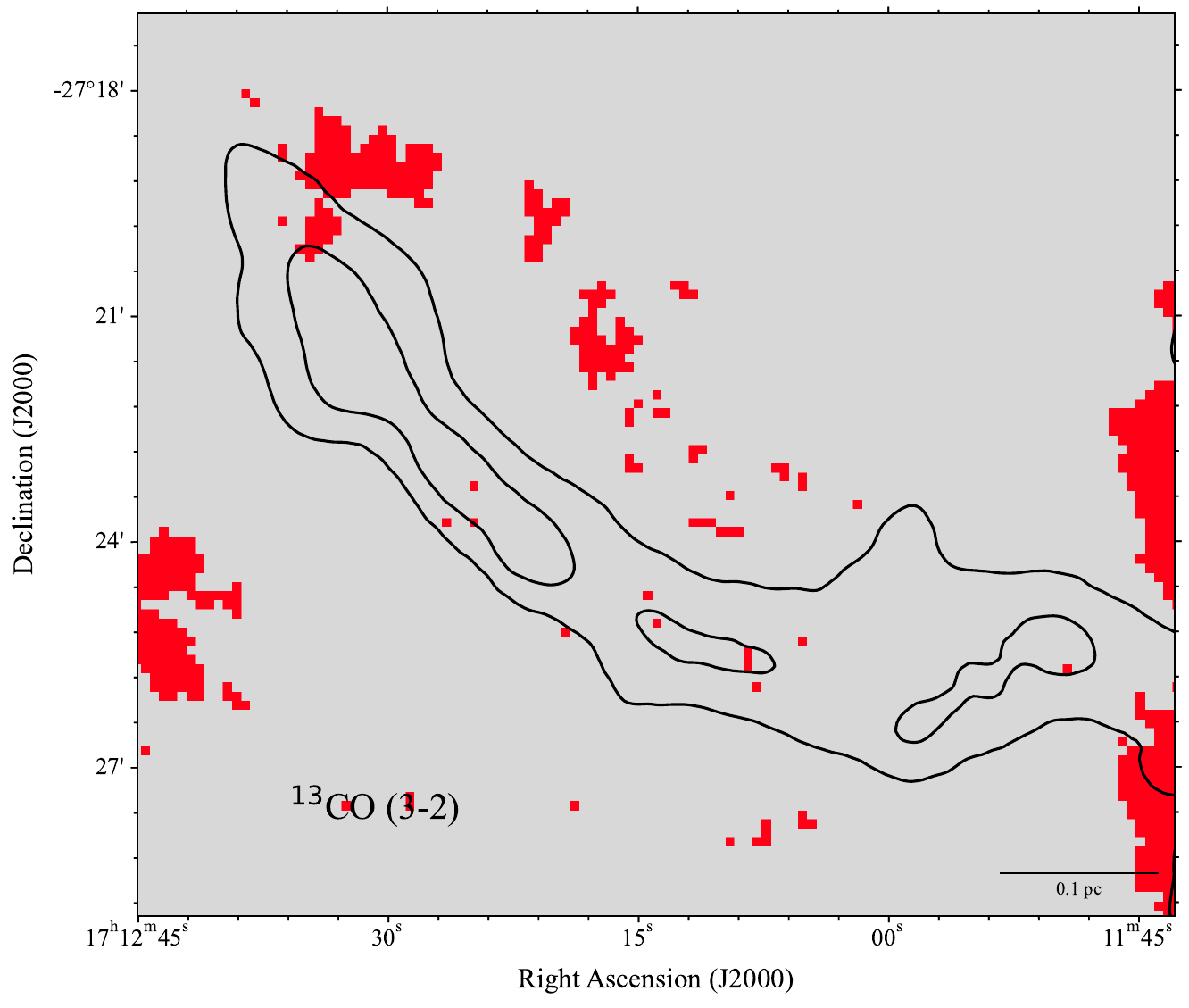} }}
    \caption{Positions where the Gaussian fit fails for \co (3-2) and \cco (3-2), shown in the top and bottom panels, respectively. Contours shows the column density of \hh with levels [4,6] $\times 10^{22}$ \cm. Scale bars are at the bottom right of each panel.}
    \label{residuals}
\end{figure}

We fit a single Gaussian component to the spectra in the \co (3-2) and \cco (3-2), using the Python $ \textsc{pyspeckit}$ library ( \citealt{ginsburg11}). All spectra that have a peak signal-to-noise ratio lower than 3.5 are excluded from the analysis for the case of \co (3-2) and lower than 4.5 (3-2) for \cco. Figure \ref{residuals} shows the positions with bad fit for \co and \cco emission lines. Red pixels display the places where the single Gaussian fit fails. We define them as places that have a residual higher than 2 $\times$ rms, and that Gaussian fits with a width broader than 0.35 \kms for both lines.

\section{Dust temperature map }
\label{app-b}

\begin{figure}
    \centering
    \includegraphics[width=1\linewidth]{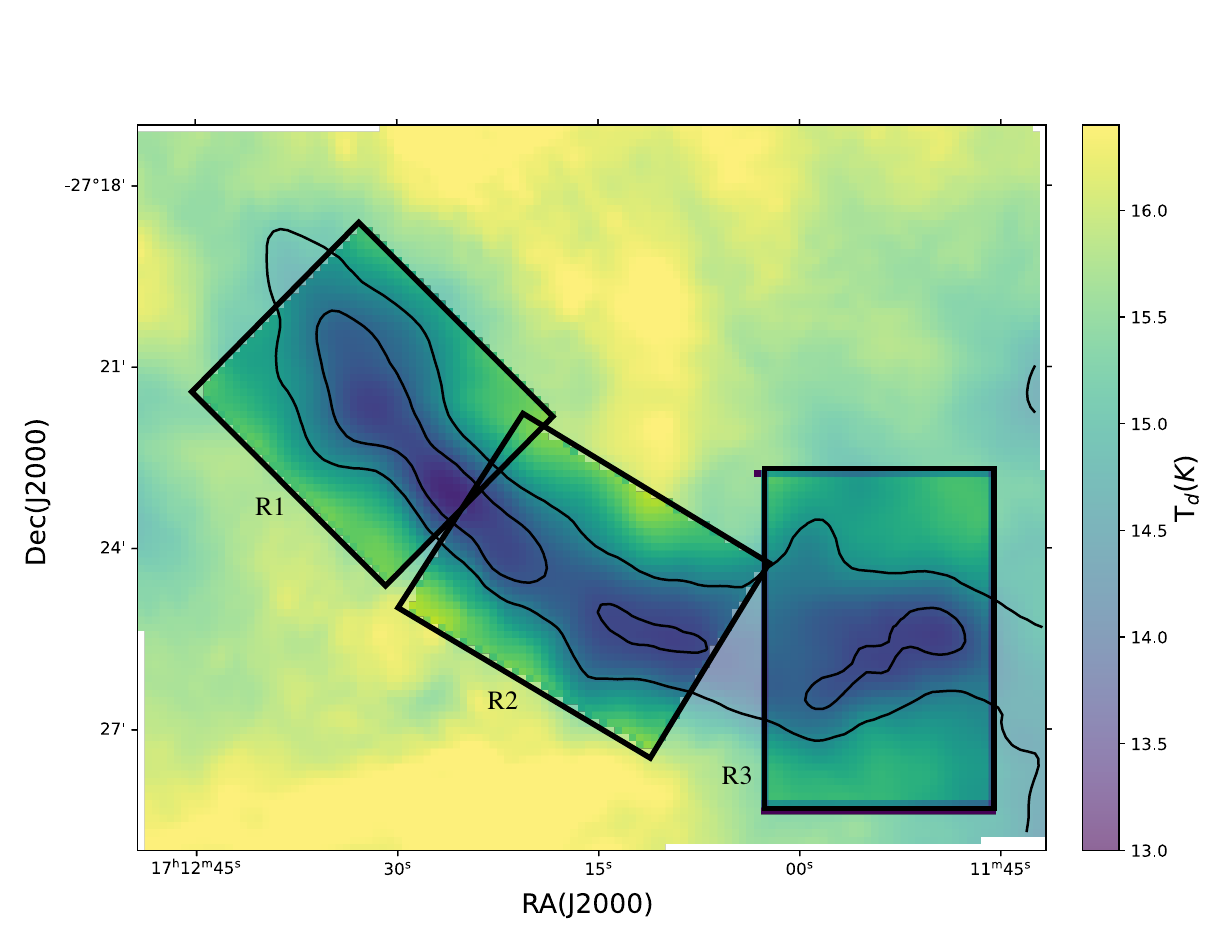}
    \caption{Dust temperature map of the filament. The black rectangles display three regions of the filament. Contours show the column density of \hh with levels [4,6] $\times 10^{22}$ \cm.  }
    \label{appen-b}
\end{figure}

Figure \ref{appen-b} represents the dust temperature map of the filament from $\emph{Herschel}$ data. The three regions (R1, R2, and R3) used in the analysis are marked by black rectangles.

\end{appendix}

\end{document}